\pdfoutput=1
\documentclass[structabstract]{aa}  
\newcommand\ergcmsa{erg\,cm$^{-2}$\,s$^{-1}$\,\AA$^{-1}$\xspace}%
\newcommand\ergcms{erg\,cm$^{-2}$\,s$^{-1}$\xspace}%

\usepackage{xspace}
\usepackage{url}

\usepackage{graphicx}
\usepackage{txfonts}
\usepackage{natbib}
\bibpunct{(}{)}{;}{a}{}{,} 
\usepackage{color} 

\usepackage{tabularx}

\begin{document}
   \title{Spectroscopy of high energy BL\,Lac objects\\with X-shooter on the VLT\thanks{Based on observations collected at the European Organisation for Astronomical Research in the Southern Hemisphere, Chile, under programs P086.B-0.135(A) and P088.B-0485(A). The raw FITS data files are available in the ESO archive.}}

   \author{S.~Pita\inst{1}\fnmsep\thanks{santiago.pita@apc.univ-paris7.fr} \and
     P.~Goldoni\inst{1} \and
     C.~Boisson\inst{2} \and
     J.-P.~Lenain\inst{3} \and
     M.~Punch\inst{1} \and           
     L.~G\'erard\inst{1}\fnmsep\inst{4} \and
     F. Hammer\inst{5} \and
     L. Kaper\inst{6} \and
     H. Sol\inst{2}
   }

   \institute{APC, AstroParticule et Cosmologie, Universit\'{e} Paris Diderot, CNRS/IN2P3, CEA/Irfu, Observatoire de Paris, Sorbonne Paris Cit\'{e}, 10, rue Alice Domon et L\'{e}onie Duquet, 75205 Paris Cedex 13, France
     \and LUTH, Observatoire de Paris, CNRS, Universit\'e Paris Diderot, 5 Place Jules Janssen, 92190 Meudon, France
     \and LPNHE, Universit\'e Pierre et Marie Curie Paris 6, Universit\'e Denis Diderot Paris 7, CNRS/IN2P3, 4 Place Jussieu, F-75252, Paris Cedex 5, France
     \and DESY, Platanenallee 6, 15738 Zeuthen, Germany
     \and GEPI, Observatoire de Paris, CNRS, Universit\'e Paris Diderot, 5 Place Jules Janssen, 92190 Meudon, France
     \and Astronomical Institute Anton Pannekoek, University of Amsterdam, Science Park 904, 1098 XH Amsterdam, The Netherlands
   }


 
\abstract
{BL\,Lac objects detected in $\gamma$-rays and, particularly, those detected at very high energies ($E>100$ GeV) by Cherenkov telescopes are extreme sources with most having redshifts lower than 0.2. Their study gives insights on the acceleration mechanisms in play in such systems and is also a valuable tool for putting constraints on the density of extragalactic background light, especially if several objects are detected at different redshifts. As their spectra are dominated by the non-thermal emission of the jet and the spectral features are weak and narrow in the optical domain, measuring their redshift is challenging. However, such a measure is  fundamental as it allows a firm determination of the distance and luminosity of the source, and, therefore, a consistent model of its emission.}
{Measurement of the redshift of BL\,Lac objects detected in $\gamma$-rays and determination of global properties of their host galaxies is the aim of this study.}
{We observed a sample of eight BL\,Lac objects with the X-shooter spectrograph installed at the ESO Very Large Telescope (VLT) to take advantage of its unprecedented wavelength coverage and its resolution, which is about 5 times higher than generally used in such studies. We extracted UVB to NIR spectra that we then corrected for telluric absorption and calibrated in flux. We systematically searched for spectral features. When possible, we determined the contribution of the host galaxy to the overall emission.}
{Of the eight BL\,Lac sources, we measured the redshift of five of them and determined lower limits for two through the detection of intervening systems. All seven of these objects have redshifts greater than 0.2. For the remaining one, we estimated, using an indirect method, that its redshift is greater than 0.175.
In two cases, we refuted redshift values reported in other publications. Through careful modelling, we determined the magnitude of the host galaxies. In two cases, the detection of emission lines allowed to provide hints on the overall properties of the gas in the host galaxies. Even though we warn that we are dealing with a very small sample, we remark that the redshift determination efficiency of our campaign is higher than for previous campaigns. We argue that it is mainly the result of the comparatively higher resolution of X-shooter.}
{}


   \maketitle

\section{Introduction}

BL\,Lac objects are active galactic nuclei (AGN), whose optical spectrum is dominated by continuum radiation emitted by the jet launched by the supermassive black hole residing in the nucleus of the galaxy. Along with flat spectrum radio quasars (FSRQs), they are the most numerous $\gamma$-ray emitters in the 0.1--300 GeV range explored by the \textit{Fermi} satellite with about 475 associations in the last Fermi AGN catalogue \citep[2LAC][]{Ackermann_2011:2LAC, Shaw_2013:zbllacs}. Their $\gamma$-ray emission is caused by particle acceleration in the jet, which has Lorentz factors $\Gamma$ up to $\sim$40 \citep[see e.g.][]{Tavecchio_2010:TeVBLL}. BL\,Lac objects are characterised by rapid variability in all energy ranges.

Broadband (from radio to $\gamma$-rays continuum) spectral energy distributions (SED) of BL\,Lac objects consist of two distinct, broad components. The low-energy component, peaking in the IR to X-rays band, is generally understood as the synchrotron radiation of a population of accelerated electrons. The high-energy one, peaking in the MeV to TeV band, in leptonic models is associated with the inverse-Compton scattering of the same electrons on the synchrotron photons \citep[synchrotron-self-Compton approaches, SSC, e.g.][]{Ginzburg_1965:Sync,Ghisellini_1989:BLL} or on ambient photons \citep[external Compton approaches, EC, e.g.][]{Sikora_1985:AGN}. 
Even if the level of contribution of hadronic processes needs to be clarified, the leptonic models seem to be favoured by the correlation often observed between the flux variations in X-rays and at very high energy (VHE, $E>100$ GeV), and by the rapid variability observed during some flares \citep[see ][]{Abramowski_2012:MWL2155}. While the one-zone SSC approach is often used quite successfully to describe the SED of most of the VHE BL\,Lac objects, recent MWL studies have highlighted the limitations of this approach in some cases \citep{Fortin_2010:APLib,Abramowski_2011:PKS2005,Wouters_2012:PKS0301,HESS_2013:PKS0301,Abramowski_2012:1ES0414,Aliu_2012:RBS0413}. The understanding of these sources, both individually and as a coherent population, requires MWL observations, allowing the description over large parts of the electromagnetic spectrum of both the shape of their SEDs and their temporal variability. 

One particular characteristic of BL\,Lac objects is the difficulty of measuring their redshift even though they are quite bright in the optical domain. Indeed, according to the usual definition of BL\,Lac objects (\citet{UrryPadovani_1995}, see also ~\citet{Ghisellini_2011} for a recent discussion), only weak emission/absorption lines with an equivalent width (EW) that is smaller than 5\,\AA\ are detected in their optical spectrum. The detection of these lines is therefore challenging and it becomes more so as the redshift increases, but it is fundamental as it allows a firm determination of the distance and luminosity of the source which in turn allows a consistent model of its emission. Moreover, the properties of the optical features give insights on the properties of the host galaxy and of its central black hole. If no intrinsic spectral feature is detected, a lower limit on the redshift may be set by the detection of an absorption system in the line of sight towards the source. In addition, the BL\,Lac object's redshift can be roughly estimated under the assumption that the host galaxy is a standard candle \citep{Sbaruf_2005:imared}. The non-detection of the galaxy's absorption features then allows a rough lower limit to be set on the redshift of the source. 

Recently, \citet{Shaw_2013:zbllacs} published a very extensive study of a sample of 297 BL\,Lac objects from the \textit{Fermi} 2LAC catalogue with unknown redshifts. They performed low resolution ({$\mathcal R$} = $\lambda / \Delta \lambda$ $\sim$ 500 -- 1000) moderately deep (1800\,seconds) spectroscopic observations using mainly three to five meter class telescopes. They measured 102 new spectroscopic redshifts (34\% efficiency of redshift determination) and 75 lower limits from foreground absorbers (25\% of the sample) for a total of 59\% sources with spectroscopic redshift or lower limits. The median of the spectroscopic redshifts is 0.33 and the median of the absorption limits is 0.70. They also estimated the redshifts from the standard candle assumption for the whole sample; however, 80\% of the time these are less constraining than the limits from line of sight absorbers.

Since the 1990's, a small fraction of BL\,Lac objects has been detected by atmospheric Cherenkov telescopes in the VHE range. The origin and properties of the VHE emission of these objects are being actively investigated as they are proof of particle acceleration at energies rarely seen in other classes of objects, demonstrating extreme acceleration properties. Beyond the study of VHE BL\,Lac objects, their detection at different redshifts is also a valuable tool to put constraints on the density of the extragalactic background light~\citep[EBL, see][]{HauserDwek_2001:EBL} because absorption is due to pair production in the interactions between VHE and EBL photons. This EBL radiation includes the UV-optical emission of all the stars and galaxies since the end of the cosmic dark ages and its reprocessing by dust in the near infrared. It therefore carries valuable information about the evolution of matter in the Universe, and also plays a role as an absorber for $\gamma$-rays. From the above considerations we can easily understand that the precise measurement of the redshifts of BL\,Lac objects and thus of their luminosities is a crucial input for the modelling of the interactions between VHE and EBL photons.

In the last decade, these studies have received a decisive boost with the advent of the third generation of atmospheric Cherenkov telescopes (H.E.S.S., MAGIC and VERITAS) which has increased the population of active galactic nuclei detected in VHE $\gamma$-rays up to 55 sources, of which 49 are BL\,Lac objects. This population is likely to be biased towards low redshifts due to the strategy of observations, which favours nearby sources to optimise rapid detection because the flux naturally decreases with distance and the $\gamma$-ray absorption by the EBL increases with increasing distance. Thus, the VHE BL\,Lac objects detected at redshifts higher than 0.2 are quite rare; only eight are known. Of these, three have measured redshifts, while five have lower limits from intervening systems \citep[][and references therein]{Halpern_1991, Rines_2003, Pita_Gamma2012:XSH, Furniss_2013}. Only two of these objects, KUV\,00311-1938 and PKS\,1424+240 have redshifts greater than 0.5. For population studies, dedicated measurement of the EBL spectrum, and tests of alternative models for VHE $\gamma$-ray production, the number of VHE BL\,Lac objects with known redshift at $z \ge 0.2$ should increase. This will depend on the development of new, more powerful facilities, such as H.E.S.S.-2 \citep{Vincent_2005:HESS2} and especially CTA \citep{CTA_2013:Concept} but also on campaigns dedicated to the determination of redshifts for BL\,Lac objects which are likely to emit in VHE.

\begin{figure*}
\centering
 \includegraphics[width=12cm]{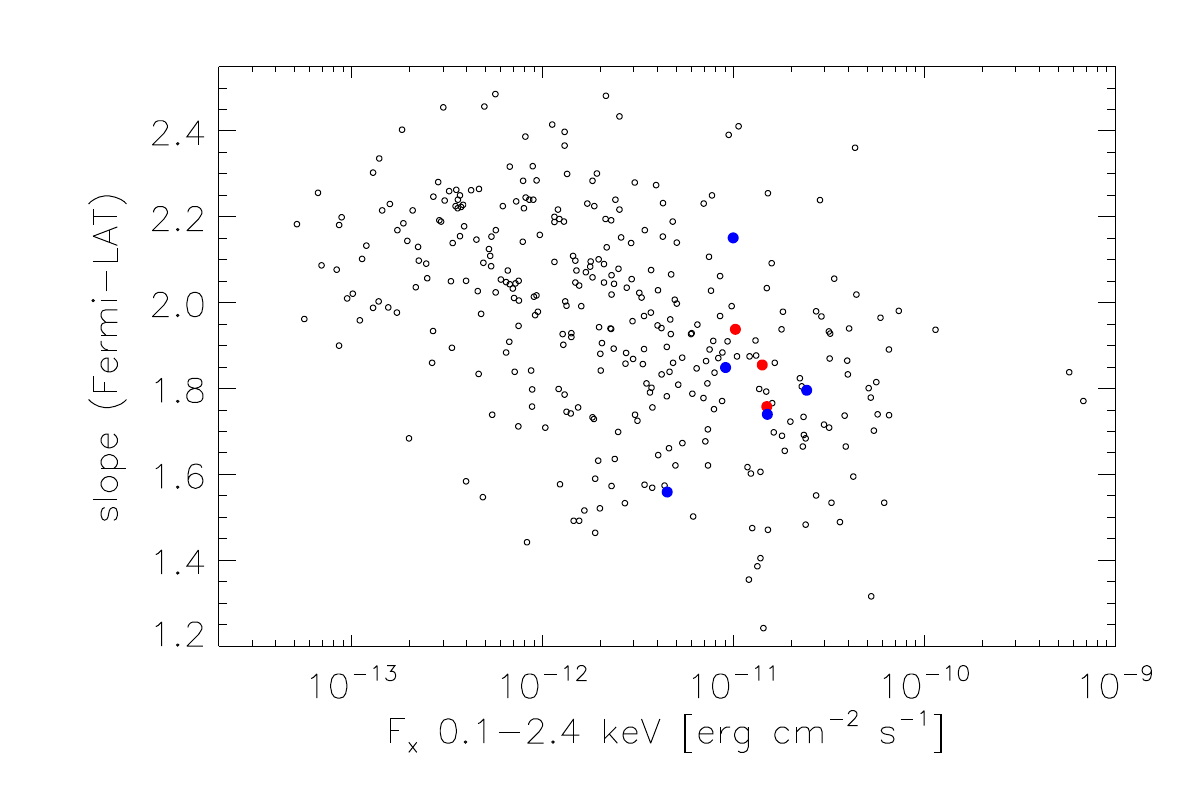}
  \caption{Distribution of the X-ray fluxes between 0.1 and 2.4 keV and the \textit{Fermi}/LAT photon indices above 100 MeV for the BL\,Lac objects population observed by \textit{Fermi}/LAT \citep{Ackermann_2011:2LAC}. Among the eight sources selected in this paper, those already detected in VHE are in red, the five others are in blue.}
  \label{FigFxGamma}
\end{figure*}

To contribute to this goal, we took advantage of the X-shooter spectrograph \citep{Vernet_2011:XSH}, which was recently installed at the VLT, which has a higher spectral resolution and a wider wavelength range than the spectrographs previously used in such searches; its sensitivity is comparable to the best of them. We selected a sample of eight sources. Three of them, namely PKS\,0447--439, KUV\,00311--1938, and PKS\,0301--243, are of special interest as they have recently been discovered as VHE $\gamma$-ray emitters with H.E.S.S., while their redshift is still unknown or needs to be confirmed. The five other BL\,Lac sources have unknown redshifts and were selected based on the following criteria: X-ray fluxes indicating the presence of high energy electrons in the jet \citep{Costamante_2002:FrFx}; spectral properties at high energy (HE,  $0.1<E<100$ GeV) on the \textit{Fermi}/LAT 2LAC catalogue \citep{Ackermann_2011:2LAC}, which suggest that they may also be detectable in the VHE range; and a good visibility during our observations. The sources selected that take these criteria into account are listed in Table~\ref{ztab}. Their X-ray fluxes and \textit{Fermi}/LAT photon indices are shown in Fig.~\ref{FigFxGamma}. The sub-set of 3 already selected VHE BL\,Lac objects is included both in Table~\ref{ztab} and Fig.~\ref{FigFxGamma} for comparison. We note that five\footnote{namely KUV\,00311--1938, BZB\,J0238--3116, BZB\,J0543--5532, BZB\,J0505+0415 and BZB\,J0816--1311, see Section~\ref{Results} for details.} of these eight sources have been observed by \citet{Shaw_2013:zbllacs}.

\begin{table*}
\caption{\label{ztab}Observed sources and selection criteria.}
\centering
\begin{tabular}{lcrcll}
\hline\hline
Source         & $F_{X}$              & $F_\mathrm{Fermi}$                          & $\Gamma_\mathrm{Fermi}$ & VHE status & $z$ status \\
               & [\ergcms]           & [$10^{-10}$ $\gamma$ $cm^{-2}s^{-1}$] &                &            &          \\
\hline
PKS\,0447--439   & $1.41 \times 10^{-11}$ & 114.0$\pm$4.0 & 1.855$\pm$0.023 & Detected & Uncertain \\
KUV\,00311--1938 & $1.49 \times 10^{-11}$ & 31.2$\pm$2.3 & 1.758$\pm$0.049 & Detected & Uncertain \\
PKS\,0301--243   & $1.02 \times 10^{-11}$ & 67.3$\pm$3.2 & 1.938$\pm$0.031 & Detected & To be confirmed \\
\hline
BZB\,J0238--3116 & $9.06 \times 10^{-12}$ & 10.0$\pm$1.5 & 1.849$\pm$0.109 & Candidate & Recently determined \\
BZB\,J0543--5532 & $1.50 \times 10^{-11}$ & 19.4$\pm$2.0 & 1.740$\pm$0.077 & Candidate & Unknown \\
BZB\,J0505+0415 & $9.93 \times 10^{-12}$ &  7.6$\pm$1.6 & 2.151$\pm$0.151 & Candidate & Uncertain \\
BZB\,J0816--1311 & $2.41 \times 10^{-11}$ & 27.5$\pm$2.3 & 1.796$\pm$0.061 & Candidate & Uncertain\\
RBS\,334        & $4.48 \times 10^{-12}$ & 3.6$\pm$1.1 & 1.559$\pm$0.217 & Candidate & Unknown\\
\hline
\end{tabular}
\tablefoot{The columns contain from left to right: (1) source name, (2) X-ray flux between 0.1 and 2.4 keV, (3) high-energy flux between 1 and 100\,GeV, (4) \textit{Fermi}/LAT photon index, (5) status of a VHE detection, and (6) redshift status. All values come from \url{http://www.asdc.asi.it/fermi2lac} (errors not available for X-ray fluxes). The sources labelled as ``Candidate'' in column 5 are those considered to be likely to be detected in VHE in a reasonable amount of time (see text for details) Explanations about the status of the sources redshifts are given in Section~\ref{Results}.}
\end{table*}

The plan of the paper is as follows. In Section~\ref{Observations}, we discuss the observations and the general methods of the data reduction. The methods to estimate the blazar emission are presented in Section~\ref{FitModel}. The observational history and our detailed results for each source are presented in Section~\ref{Results}. We then discuss the results of our observations in the context of high redshift blazars in Section~\ref{Discussion} and conclude in Section~\ref{Conclusion}. For all calculations, we used a cosmology with $\Omega_{\rm M}$ = 0.27, $\Omega_{\rm \Lambda}$ = 0.73 and H$_0$ = 71 km s$^{-1}$ Mpc$^{-1}$. All wavelengths in the paper are in vacuum.

\section{Observations and data reduction}
\label{Observations}
The X-shooter spectrograph \citep{Vernet_2011:XSH} is a single-object medium resolution ({$\mathcal R$} = $\lambda$/$\Delta$$\lambda$ = 5000 -- 10000) {\'e}chelle spectrograph that started operations in 2009 October on the VLT. It is the first second-generation VLT instrument and was developed by a consortium comprising institutes from Denmark, France, The Netherlands, Italy and ESO. The main characteristic of X-shooter is its unprecedented simultaneous wavelength coverage from 3000\,\AA\ to 24000\,\AA. This is obtained by splitting the light using dichroics in three arms: UVB ($\lambda$ = 3000 -- 5600\,\AA), VIS ($\lambda$ = 5500 -- 10200\,\AA), and NIR ($\lambda$ = 10000 -- 24000\,\AA). Its resolution {$\mathcal R$} is between 3000 and 17000 depending on arm and slit width. For these observations, we choose slit widths of 1.3\arcsec, 1.2\arcsec\ and 1.2\arcsec\ for the UVB, VIS, and NIR arms, which results in {$\mathcal R$} $\sim$ 4000, 6700, and 3900 respectively. X-shooter is very well suited to the determination of redshift in bright blazars: on the one hand, the wavelength range allows one to search for multiple features in a very broad spectral domain; on the other hand the resolution is sufficient to detect narrow absorption lines from intervening clouds and broad absorption or emission lines from the host galaxy itself. Its high sensitivity allows us to push the search for lines to very constraining upper limits. The observing time was obtained under the framework of the French X-shooter GTO program.

\begin{table*}
\caption{\label{tabobs}Sample and parameters of the observations along with the range of estimated S/N for each arm.}
\centering
\begin{tabular}{lccccccccc}
\hline\hline
Name & RA & Dec  & Start Time & Time  & Airmass & Seeing       & S/N & S/N & S/N  \\
     &    &      &   UTC      &   (sec)  &        & (\arcsec)    & (UVB) & (VIS) & (NIR)\\  
\hline
RBS\,334 & 02 37 34.0 & -36 03 32.4 & 2010-11-12 00:33:40 & 4800 & 1.37 & 0.96 & \hspace{1.5mm}50 -- 120 & 40 -- 80 & 20 -- 50 \\
BZB\,J0238--3116 & 02 38 32.5  & -31 16 58.5 & 2011-11-30 01:59:46 & 3600 & 1.02 & 0.81 & \hspace{1.5mm}50 -- 120 & 40 -- 80 & 25 -- 50 \\
PKS\,0447--439 & 04 49 24.7 & -43 50 09.0 & 2011-11-30 04:35:34 & 4800 & 1.06 & 0.76 & 200 -- 400 & \hspace{1.5mm}60 -- 150 & 45 -- 85\\
BZB\,J0816--1311 & 08 16 27.2 & -13 11 53.0 & 2011-11-30 06:29:08 & 6000 & 1.07 & 0.80 & \hspace{1.5mm}45 -- 110 & 40 -- 55 & 20 -- 25 \\
KUV\,00311--1938  & 00 33 34.3 & -19 21 34.0 & 2011-12-01 00:32:40 & 7200 & 1.05 & 0.99 & 100 -- 250 & \hspace{1.5mm}80 -- 200 & 25 -- 80\\
PKS\,0301--243 & 03 03 26.5 & -24 07 11.5 & 2011-12-01 04:11:45 & 4800 & 1.02 & 1.15 & \hspace{1.5mm}70 -- 210 & \hspace{1.5mm}75 -- 140 & 35 -- 70\\
BZB\,J0505+0415 & 05 05 34.8 & +04 15 54.6 & 2011-12-01 05:06:17 & 4800 & 1.16 & 1.16 & 40 -- 90 & 50 -- 80 & 20 -- 35\\
BZB\,J0543--5532 & 05 43 57.3 & -55 32 08.0 & 2011-12-01 06:56:59 & 4800 & 1.24 & 1.03 & \hspace{1.5mm}50 -- 110 & 65 -- 95 & 20 -- 60\\ 
\hline
\end{tabular}
\tablefoot{The columns contain from left to right: (1) source name, (2) right ascension (J2000), (3) declination (J2000), (4) start time of the observations, (5) exposure time, (6) average airmass, (7) average seeing at Zenith in the $R$ band, and (8, 9, 10) S/N ranges measured in selected regions of the continuum in the UVB, VIS, and NIR arms, where no obvious features were present. The slit widths are the same for all observations: 1.3\arcsec, 1.2\arcsec\ and 1.2\arcsec\ for the UVB, VIS, and NIR arm, respectively. }
\end{table*}

The observations took place on 2011 November 30 for all the sources, except for one of them, which was observed on 2010 November 12 (see details in Table \ref{tabobs}). The exposures were taken using the “nodding along the slit” technique with an offset of  5\arcsec\ between exposures of 600 seconds each, usually in a standard ABBA sequence. Each observation was preceded or followed by an observation of an A0V telluric standard star at similar airmass.

We processed the spectra using  version 1.3 of the X-shooter data reduction pipeline \citep{gol06,mod10}. The raw frames were first subtracted and cosmic-ray hits were detected and corrected using the method developed by \citet{vdok01}. The frames were divided by a master flat field obtained by using day-time flat field exposures with halogen lamps. The orders were extracted and rectified in wavelength space using a wavelength solution previously obtained from calibration frames.

The resulting rectified orders frames were then shifted and added to superpose them, thus obtaining the final 2D spectrum. The orders were then merged, and, in the overlapping regions, the merging was weighted by the errors, which were propagated during the process. 

From the resulting 2D merged spectrum of all sources, we extracted a one dimensional spectrum with a corresponding error file and a bad pixel map at the source's position. We checked every one of these spectra for obvious spectral features that could allow the measurements of the redshift. For those spectra that showed no features, we also extracted three sub-spectra from three sub-apertures with one centred on the source position and the others on the opposite sides. The width of these sub-apertures were adjusted for each source depending on its size in the slit direction. These lateral spectra allowed us to explore the emission in regions less affected by the glare of the blazar.

At the end of our first reduction, we noticed that the flux levels of overlapping orders were not always compatible in the UVB and VIS spectra. As a consequence, the merged spectrum showed jumps of a few percent, which are well visible at our high signal-to-noise levels. This is a well-known effect in {\'e}chelle spectroscopy for high signal-to-noise spectra that has been possibly attributed to smooth, time dependent changes in the light path \citep{Hensberge_2007:RemoveInstrumental}. As this effect was limited only to the final (initial) part of the orders, we did not extract the final part of the non-compatible orders. We thus produced merged spectra with smooth transitions between orders while lowering signal-to-noise only by a few percent.

\subsection{Flux Calibration and telluric correction}

Absolute flux calibration and telluric correction of {\'e}chelle spectra are not straightforward processes, and the exceptional wavelength coverage of X-shooter complicates the matter further. However, these two tasks are very important for our goals. On the one hand we want to reconstruct the general physical shape of the spectrum which allows us to estimate the contribution of the jet and of the galaxy to the overall emission. On the other hand, telluric corrections are mandatory, as spectral features may be hidden by the atmospheric absorption.

We tried different approaches to achieve our goals. First, we produced flux-calibrated spectra using the pipeline by producing a response function with wide slit observations of the flux standard Feige 110 obtained on the same night. The reduction of the flux standard was performed as above, but we subtracted the sky emission lines using the method of \citet{kel03} in this case. The extracted spectrum was divided by the flux table of the star from the calibration table available within the pipeline \citep{Vernet_2010:Standard} to produce the response function, which was then applied to the spectrum of the science target during its reduction.

These pipeline-calibrated spectra allowed the general spectral shape to be reproduced; however, small instrumental features, due to the coarse rebinning of the response function, were still present therefore precluding the search of spectral features at some wavelengths. We then produced a response function dividing bin by bin the spectrum of the star by the HST\footnote{The Hubble Space Telescope.} flux table of Feige 110 from the CALSPEC database~\citep{boh07}, which we interpolated on the wavelength grid of our spectrum. We then applied a median filter to avoid the strongest instrumental features and used this new response function to calibrate in flux the spectra of the science targets. Finally, we corrected the result from slit losses, which were estimated with public X-shooter observations of Feige 110 in narrow and wide slit mode obtained a few nights later. This estimation was computed by taking into account the different airmasses of the standard star observations. In the following, we call these spectra flux-calibrated spectra.

We used the A0V telluric standard stars to correct for the telluric absorption in the VIS and NIR spectra of the BL\,Lac objets. The telluric standard spectra were extracted with the same procedure used for the flux standard, and we used these spectra to apply telluric corrections and flux calibrations simultaneously with the package SpeXtool~\citep{vac03}. The flux scale was set using the $B$ and $V$ magnitudes of the telluric stars as available in Simbad\footnote{\url{http://simbad.u-strasbg.fr/simbad/}}. In this case we note that no slit loss correction is needed as the slit width is the same for the science target and the telluric standard. We refer to these spectra as telluric-calibrated spectra.

The flux levels of the flux-calibrated and telluric-calibrated spectra are broadly consistent, but their spectral slopes are slightly different. The average difference between the spectra is about 20 \% at the blue end of the VIS spectra and tends to zero at redder end of the NIR spectra. This is consistent with the presence of residual slit losses in the flux-calibrated spectra. 

We then decided to use the VIS and NIR telluric-calibrated spectra and the UVB flux-calibrated spectra for our further analysis. In case of discrepancies in the absolute flux, which were at most of the order of a few percent, we adjusted the flux of the VIS and NIR spectra to the flux of the UVB spectrum. 

Finally, we corrected the spectra for Galactic extinction using the maps of \citet{Schle98} and the wavelength dependent extinction curves of \citet{Fitz99}.

\begin{table*}[th]
\caption{\label{magtab}Spectrophotometric magnitudes.}
\centering
\begin{tabular}{lccccccccccc}
\hline\hline
Source & \multicolumn{8}{c}{Filters} & \multicolumn{3}{c}{2MASS}\\
         &  U  &  B    &  V   &  Rc  &  Ic    & J & H & Ks & J & H & K\\ 
\hline
PKS\,0447--439 & 13.4   &  14.2  & 13.8 & 13.5  & 13.1  & 12.4  & 11.6 & 10.9 & 13.9 & 13.2 & 12.6\\
PKS\,0301--243 & 14.7   & 15.5   & 15.1  & 14.7  & 14.3   & 13.4  & 12.6 & 11.7 & 14.5 & 13.7 & 13.1\\
KUV\,00311--1938 &  16.0  & 16.7   & 16.4   & 16.1   & 15.7  & 15.0   & 14.2  & 13.3 & 14.7 & 14.1 & 13.3\\
BZB\,J0238--3116  & 16.3  & 17.1 & 16.8 & 16.4 & 16.0 & 15.3 & 14.6 & 13.8 & 15.2 & 14.4 & 13.9\\
BZB\,J0543--5532 & 16.3 & 17.1 & 16.8 & 16.4 & 16.1 & 15.5 & 14.8 & 14.1 & 15.1 & 14.5 & 13.8\\ 
BZB\,J0505+0415 & 16.8 & 17.5 & 17.2 & 16.8 & 16.3 & 15.5 & 14.7 & 14.0 & 15.5 & 14.7 & 13.9\\
BZB\,J0816--1311 & 16.7  & 17.5  & 17.1  & 16.7 & 16.3  & 15.6 & 14.9 & 14.1 & 15.1 & 14.4 & 13.8\\
RBS\,334 & 16.7 & 17.4 & 17.1 & 16.8 & 16.4 & 15.8 & 15.1 & 14.4 & 16.0 & 15.4 & 14.4\\
\hline
\end{tabular}
\tablefoot{The spectrophotometric magnitudes are corrected for Galactic extinction of the observed sources in the UVBRI Johnson-Cousins and 2MASS JHKs filters. The error is $\pm$ 0.15 mags for all filters (see text for details). For comparison, the 2MASS J, H, and K magnitudes \citep{Skrutskie_2006:2MASS} are given in the three last columns.}
\end{table*}

\subsection{Synthetic photometry}
\label{Phot}
We estimated the magnitudes of the sources by performing synthetic photometry on our spectra. We convolved our spectra with standard imaging filters, namely the Johnson-Cousins UBVRI filters and the 2MASS JHK$_{\rm s}$ filters. The filters parameters were taken from the ADPS web site\footnote{Asiago Database on Photometric Systems, \url{http://ulisse.pd.astro.it/Astro/ADPS/Systems/index.html}}.

To test the precision of our procedure, we flux calibrated the telluric standard stars and computed their spectrophotometric magnitudes as we did for the science targets. We then compared this result to their Simbad magnitudes. The difference between archival and measured magnitudes had a dispersion of 0.1 magnitudes (1$\sigma$) in all filters, we therefore used this value as an estimate of the error in our procedure. Moreover, we found an offset of $-$0.3 $\pm$ 0.1 magnitudes between the archival and the measured magnitudes in the $U$ and $B$ filters. This offset tends to zero towards longer wavelengths. As we used the UVB arm to set the flux scale, we subtracted this offset to our measured values. The final error on the magnitude was $\sim$ $\pm$ 0.15 magnitude; this is the quadratic sum of the two estimates discussed above. The results are in Table~\ref{magtab}, along with the 2MASS survey magnitudes. During our observations, PKS\,0447--439 was about 1.5 magnitude brighter than in the 2MASS survey, while PKS\,0301--243 was about 1 magnitude brighter. All the other sources magnitudes can be found within half a magnitude from the 2MASS values.

\begin{table}[!h]
\caption{\label{TabLines} List of the strongest emission or absorption lines.}
\begin{tabular}{lll}
\hline\hline
Feature names & Wavelength (\AA) & Type \\  
\hline
Ly$\alpha$ & 1215  & Absorption/Emission\\
Fe II & 2600 & Intervening\\
Mg II & 2796 & Intervening/Emission\\
      & 2803 & Intervening/Emission\\
{[OII]} & 3727 & Emission\\
      & 3729 & Emission\\
Ca K & 3933.7 & Absorption\\
Ca H & 3968.5 & Absorption\\
H$\delta$ & 4101.7 & Absorption\\
Ca I G & 4304.4 & Absorption\\
H$\beta$ & 4861.3 & Absorption/Emission\\
{[OIII]} & 4959 & Emission\\
     & 5007 & Emission\\
Mgb & 5174 & Absorption\\
CaI+FeI & 5269 & Absorption\\
Na I D & 5892.5 & Absorption\\
{[NII]} & 6548.1  & Emission\\
H$\alpha$ & 6562.8 & Absorption/Emission\\
{[NII]} & 6583.6 & Emission\\
Ca triplet & 8498 & Absorption\\
           & 8542  & Absorption\\
           & 8662  & Absorption\\
\hline
\end{tabular}
\tablefoot{The columns give the names of the strongest expected absorption and emission lines, their wavelength in the rest frame in \AA, and their nature, which indicate if they appears as absorption or emission lines from the host galaxy, or if they come from an intervening system. For Fe II and Mg II, which are multiplets, only the strongest lines are reported.}
\end{table}

\section{Estimation of the Blazar emission}
\label{FitModel}

In the optical-ultraviolet range, the observed SED of a blazar is a combination of jet emission, AGN activity (thermal and non-thermal), and thermal emission of the host galaxy, which usually is an elliptical \citep{Urry_1999:Elliptical}. Depending on the relative strength of each component, the spectrum might be featureless, or it may display signatures coming from the host galaxy. This is clearly shown in the simulations performed by \citet{Landt_2002} and \citet{Piranomonte_2007:SedentarySurvey}. We modelled the emission using a combination of a power law (featureless continuum) with the local elliptical template of \citet{Mannucci_2001:Template} by adding Gaussian emission features when needed. The fit was performed via a $\chi^2$ minimisation using the MINUIT package\footnote{\url{http://root.cern.ch/root/html/TMinuit.html}}. 

The free parameters of the model are the power law wavelength index, its normalisation at 10000\,\AA\ and the flux level at 10000\,\AA\ of the host galaxy emission. If the redshift was previously determined by the detection of a feature, the fit was performed at that redshift, thereby allowing us to estimate the flux of the host galaxy. The statistical errors provided by the fitting method are in general unphysically small (typically $10^{3}$ times lower than the fitted value). Therefore, by following the approach of \citet{Shaw_2013:zbllacs}, we independently fitted separated sections of the spectra and estimated the error from the differences between the resulting parameters.

When no feature (or only an intervening absorption system) was detected, we fitted the spectrum with a simple power law with a normalisation at 10000\,\AA\ as above and we estimated the fitting errors in the same way. We also estimated a lower limit on the redshift with the imaging method of \citet{Sbaruf_2005:imared} using a host galaxy magnitude $M_\mathrm{R} = -22.5$ \citep{Shaw_2013:zbllacs}. Moreover, to estimate an upper limit to the galaxy contribution, we constructed a grid of blazar emission models at the redshift of the farthest absorbing system (if present) or at the redshift determined with the imaging method. These models were built as the sum of the measured power law and of a template with growing flux to which we added Poissonian noise to achieve the signal-to-noise of our observations. For each model in the grid, we measured the EW of the Ca II H\&K feature, which is the first to be clearly visible at these redshifts. The upper limit was fixed when the feature was detected at the 2 $\sigma$ level. In one case, namely for PKS\,0301--243, we used published photometry of the host galaxy to estimate the flux of the template; the upper limit to the template flux was determined as discussed above. To account for the emission lines detected in two cases, we added narrow lines with a full width at half maximum (FWHM) of 500 km/s to the template, which are compatible with central gas dispersion in local galaxies \citep{Verdoes_2006}. This allowed a rough estimation of the line ratios in these objects.

Results are given in Table~\ref{TabResults} for sources for which we determined a firm redshift and in Table~\ref{TabResultsLL} for the others.
 
\section{Sources and results}
\label{Results}

In the following, we discuss the results of our observations for each source. The spectra are modelled as described in Section~\ref{FitModel}. For the determination of the redshift, we searched for emission or absorption lines of the host galaxies and for intervening absorbers following the list in Table~\ref{TabLines}.

\subsection{PKS\,0447--439}

\begin{figure*}
\centering
 \includegraphics[width=15cm]{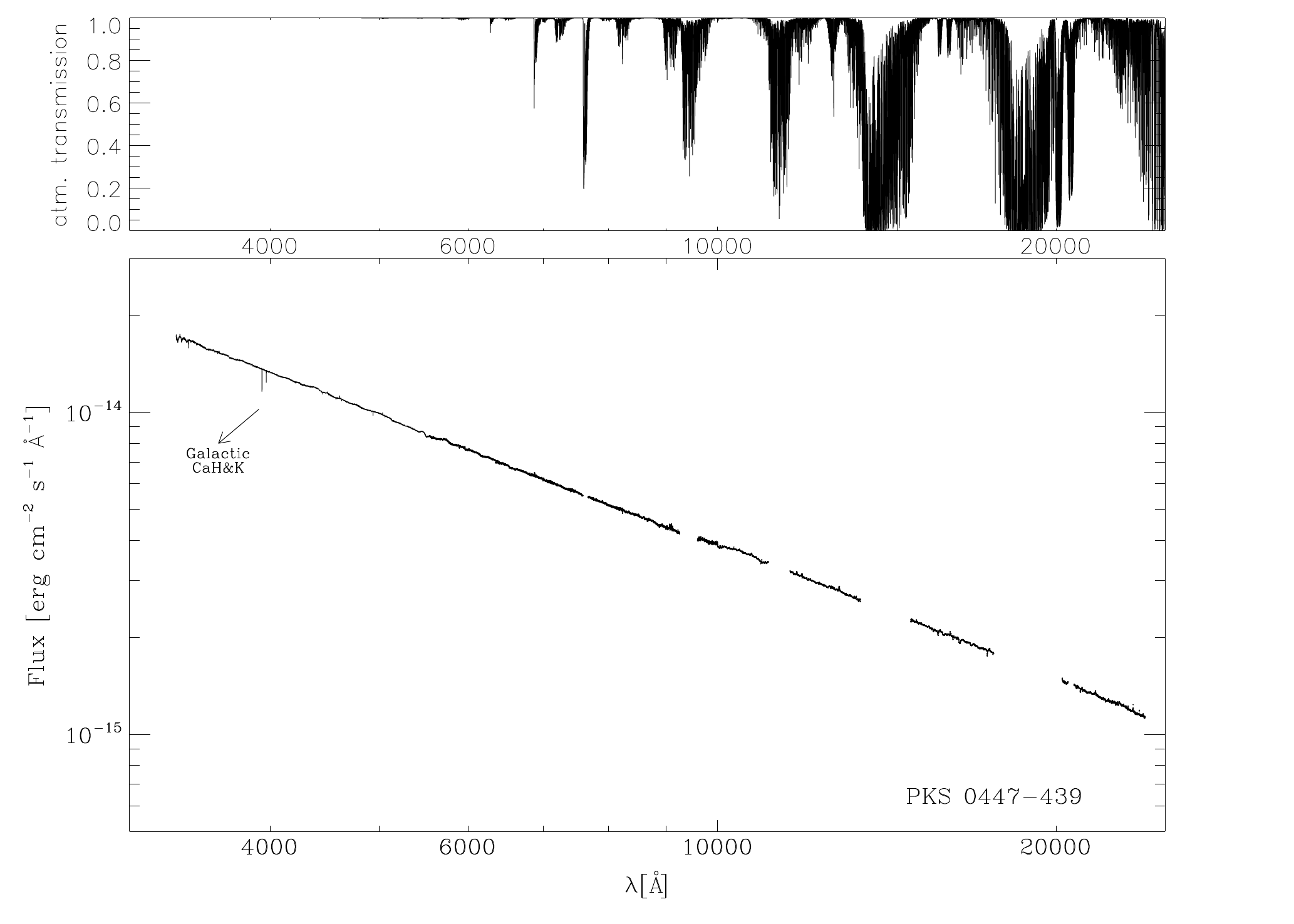}
  \caption{\label{FigPKS0447:Broad}  The UV-to-NIR spectrum of PKS\,0447--439 obtained with X-shooter after flux calibration and telluric corrections. The spectrum is dominated by the non-thermal emission from the nucleus and is well described by a power law (see Table~\ref{TabResultsLL} for details). It shows no spectral feature, except the Ca II H\&K doublet associated with our own Galaxy. The typical atmospheric transmission for Paranal is shown in the upper panel.
}
\end{figure*}

\begin{figure*}
\centering
 \includegraphics[width=15cm]{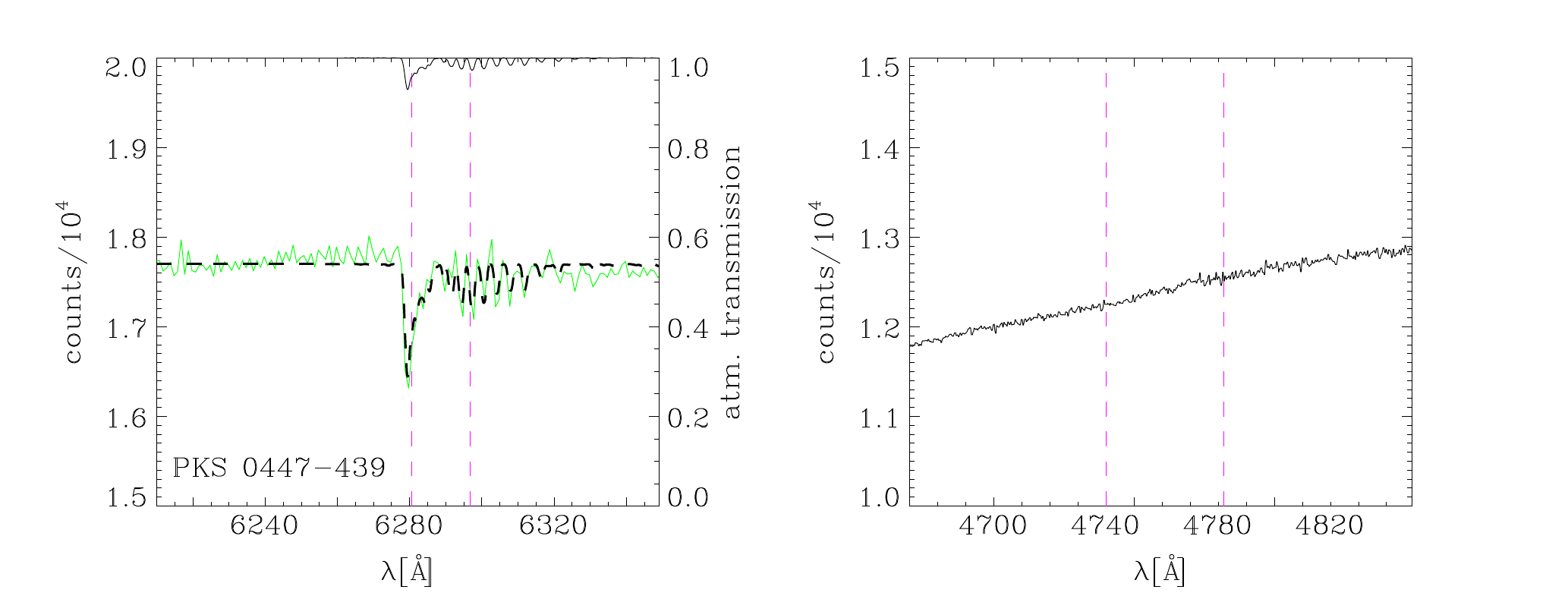}
  \caption{\label{FigPKS0447:Region} Left: The PKS\,0447--439 counts spectrum (continuous green line in the middle) in the region 
where \cite{Landt_2012} detects a feature that she interprets as the redshifted Mg II 2800\,\AA\ doublet (at a position shown by dashed vertical lines). 
The expected transmission rate, smeared to the resolution of our data, is shown in the top. Once normalised to the flux level of 
the PKS\,0447--439 we show that the expected transmission rate (dashed curve) explains the shape of the PKS\,0447--439 counts spectrum. 
Right: The PKS\,0447--439 counts spectrum in the region where \citet{Perlman_1998:PKS0447} detected the feature associated to the Ca II H\&K doublet. The position expected for the Ca II H\&K doublet at $z = 0.205$ is indicated by the vertical dashed lines.
}
\end{figure*}

The object PKS\,0447--439 is one of the brightest high energy extragalactic sources detected by \textit{Fermi}/LAT \citep{Abdo_2009:BrightAGN,Ackermann_2011:2LAC} and was recently discovered as a VHE emitter by the H.E.S.S. Cherenkov telescopes array \citep{Raue_2009:PKS0447,Zech_2011:PKS0447,Abramowski_2013:PKS0447}. Initially detected as a radio source with the Molonglo telescope \citep{Large_1981:Molonglo}, it has been since observed at different wavelengths (see \citet{Abramowski_2013:PKS0447} for a short review). After being misidentified as a Seyfert 1 galaxy by \citet{Craig_1997:OpticalSeyfert} (see \citet{Prandini_2012:PKS0447} for comments on this misidentification), the source was identified as a bright BL\,Lac object by \citet{Perlman_1998:PKS0447} and classified as an HBL by \citet{Landt_2008}, based on the ratio of the radio core luminosity at 1.4 GHz over the X-ray luminosity at 1 keV.

The redshift of this source is not yet clearly established, although several values or constraints have been proposed:
\begin{itemize}
\item $z = 0.107$, a value based on the misidentification of the source with a Seyfert 1 galaxy \citep{Craig_1997:OpticalSeyfert};
\item $z = 0.205$, a value based on a very weak spectral feature identified as the Ca II H\&K doublet in an otherwise featureless spectrum obtained with the CTIO 4m telescope \citep{Perlman_1998:PKS0447};
\item $z > 0.176$ \citep{Landt_2008}, a lower limit based on a photometric method  \citep{Piranomonte_2007:SedentarySurvey} that is applied on a featureless spectrum obtained with the CTIO and NTT telescopes;
\item $z \leq 0.51$ \citep{Zech_2011:PKS0447}, a reasonably model independent upper limit, which can be pushed down to z$<$0.3 when the assumption of a simple one-zone synchrotron self-Compton scenario accounts for the observed SED, derived from the $\gamma$-ray spectra determined with \textit{Fermi}/LAT and H.E.S.S.; 
\item $z = 0.20 \pm 0.05$ \citep{Prandini_2012:PKS0447}, an estimation based on a rough calibration of the redshift of a BL\,Lac from the measurement of the spectra with \textit{Fermi}/LAT and H.E.S.S. for a small set of BL\,Lacs with known redshifts;
\item $z > 1.246$, based on weak absorption lines at 6280\,\AA\ which are misidentified as the redshifted Mg II 2800\,\AA\ doublet in a recent analysis \citep{Landt_2012} of CTIO and NTT observations from 2007, giving an extremely high lower limit.
\end{itemize}

While the redshift proposed by \citet{Perlman_1998:PKS0447} at $z = 0.205$ needs to be confirmed because of the weakness of the spectral feature used for its determination, it appears to be consistent with the different non-spectroscopic estimations by \citet{Landt_2008,Zech_2011:PKS0447,Prandini_2012:PKS0447}.  The surprising lower limit of z$>$1.246 proposed by \citet{Landt_2012}, if confirmed, would imply that our understanding of the VHE $\gamma$-ray propagation is incorrect, or that the EBL density is far below the value obtained from galaxy counts, which seems to be in contradiction with the recent results from \citet{Ackermann_2012:EBL,HESS_2013:EBL}. Our results, shown below, clearly invalidate this lower limit.

The observation of PKS\,0447--439 with X-shooter took place in November 2011 with an exposure time of one hour that results in a spectrum with a S/N between 45 and 400, which depends on the arm (see Table~\ref{tabobs}). The near-IR to UV spectrum is strongly dominated by the emission of the nucleus, as shown in Fig.~\ref{FigPKS0447:Broad} (the typical atmospheric transmission for Paranal\footnote{see \url{http://www.eso.org/sci/facilities/eelt/science/drm/tech_data/background}} is shown in the upper panel). The spectral shape is close to a power law with an index $\alpha = -1.32 \pm 0.09$ and an absolute flux at 10000\,\AA\ of (3.87 $\pm$ 0.16)$\times$10$^{-15}$\,\ergcmsa. The source is clearly in a high state, as shown by the comparison of the corresponding NIR magnitudes (see Table~\ref{magtab}) with the 2MASS ones \citep{Skrutskie_2006:2MASS}. This high flux hampers the detection of features from the host galaxy and thus makes the spectroscopic determination of the redshift difficult. 

The search for absorption or emission lines in the whole wavelength range has been done within the 2 cycles of data recording and splitting of the data along the slit as indicated in Section~\ref{Observations}. After correction of the telluric spectrum (Section~\ref{Phot}), the spectrum is featureless, except the Ca II H\&K doublet associated with our own Galaxy. So, despite favourable resolution and signal-to-noise levels, there is no evidence of an extragalactic spectral feature, which could have allowed a direct spectral determination of the redshift of the source.

As already shown in \citet{Pita_Gamma2012:XSH}, we carefully examined the region around 6280\,\AA\ where \cite{Landt_2012} detects a feature that she interprets as the redshifted Mg II 2800\,\AA\ doublet. We clearly detect a band of molecular absorption from atmospheric $\mathrm{O_2}$ at these wavelengths, which is visible in all our observations. A zoom on this region is shown in the left panel of Fig.~\ref{FigPKS0447:Region}, where our spectrum corresponds to the (green) continuous line in the middle and the expected transmission rate, smeared to the resolution of our data, is shown in the top. We see that the observed feature and the expected absorption by $\mathrm{O_2}$, which is normalised to the flux level of the source (dashed line), match very well in both shape and wavelength position. This is strong evidence that the association of the 6280\,\AA\ feature with the Mg II doublet by \cite{Landt_2012} is not correct\footnote{This result has been confirmed with an independent measurement by \citet{Fumagalli_2012:PKS0447}}. We suggest that her misidentification may have been caused by the lower resolution of their spectrum ({$\mathcal R$} $\sim$ 1000), which smears the multiple molecular absorption of the $\mathrm{O_2}$ band into one or maybe two large absorption features (see their Fig.~2 with respect to our Fig.~\ref{FigPKS0447:Region}). We note that this feature, being due to the Earth's atmosphere, is present in all the optical spectra of bright sources, BL\,Lac objects and standard stars, obtained in this work.

We also examined the regions where  absorption and emission lines are expected by considering the redshift of z$=$0.205 proposed by \citet{Perlman_1998:PKS0447}, but no evidence of spectral features was found, in particular in the region where the feature associated to the Ca II H\&K  doublet in \citet{Perlman_1998:PKS0447} was detected, as shown in Fig.~\ref{FigPKS0447:Region}. It should be noted, however, that this region is dominated by the non-thermal emission of the nucleus, which strongly dilutes the spectral features from the AGN environment.

In addition, we estimated a lower limit on the redshift with the imaging method proposed by \citet{Sbaruf_2005:imared} using a host galaxy magnitude $M_\mathrm{R} = -22.5$ \citep{Shaw_2013:zbllacs} and $M_\mathrm{R} = -22.9$ \citep{Sbaruf_2005:imared}; we obtained corresponding values of $z > 0.175$ and $z > 0.19$, similar to the $z > 0.176$ already proposed by \citet{Landt_2008}.

Finally, it is possible to estimate an upper limit to the redshift from the absence of Lyman $\alpha$ absorption in the spectrum. In a first approximation, our wavelength coverage sets a rough redshift upper limit at $z \le 1.5$. A more precise upper limit may be set using the density of the Lyman $\alpha$ forest for $z \le 1.5$ and dN/dz $\propto -1.85$ for weak absorbers (EW $\ge$ 0.24\,\AA, see \citet{Bechtold94}). From these values, we estimated the redshift range giving a 1$\sigma$ probability of not detecting an absorber of that EW as $\Delta z \sim 0.03$. Moreover, we fixed the effective start of the spectrum, $\lambda_\mathrm{min}$, as the wavelength where the S/N becomes high enough to detect these absorbers. The limit was then computed with the following equation: $z_\mathrm{max}$ = ($\lambda_\mathrm{min}-1215)/1215$ +  $\Delta z$, obtaining $z_\mathrm{max}$ = 1.51.

In conclusion, while we invalidate the lower limit $z > 1.246$ proposed by \cite{Landt_2012} because the feature used in the estimation is not of extragalactic origin, we nonetheless constrain the redshift of the source with an estimated lower limit from photometry $z > 0.176$ and a solid spectroscopic upper-limit $z < 1.51$.

\subsection{KUV\,00311--1938}

\begin{figure*}
\centering
\includegraphics[width=15cm]{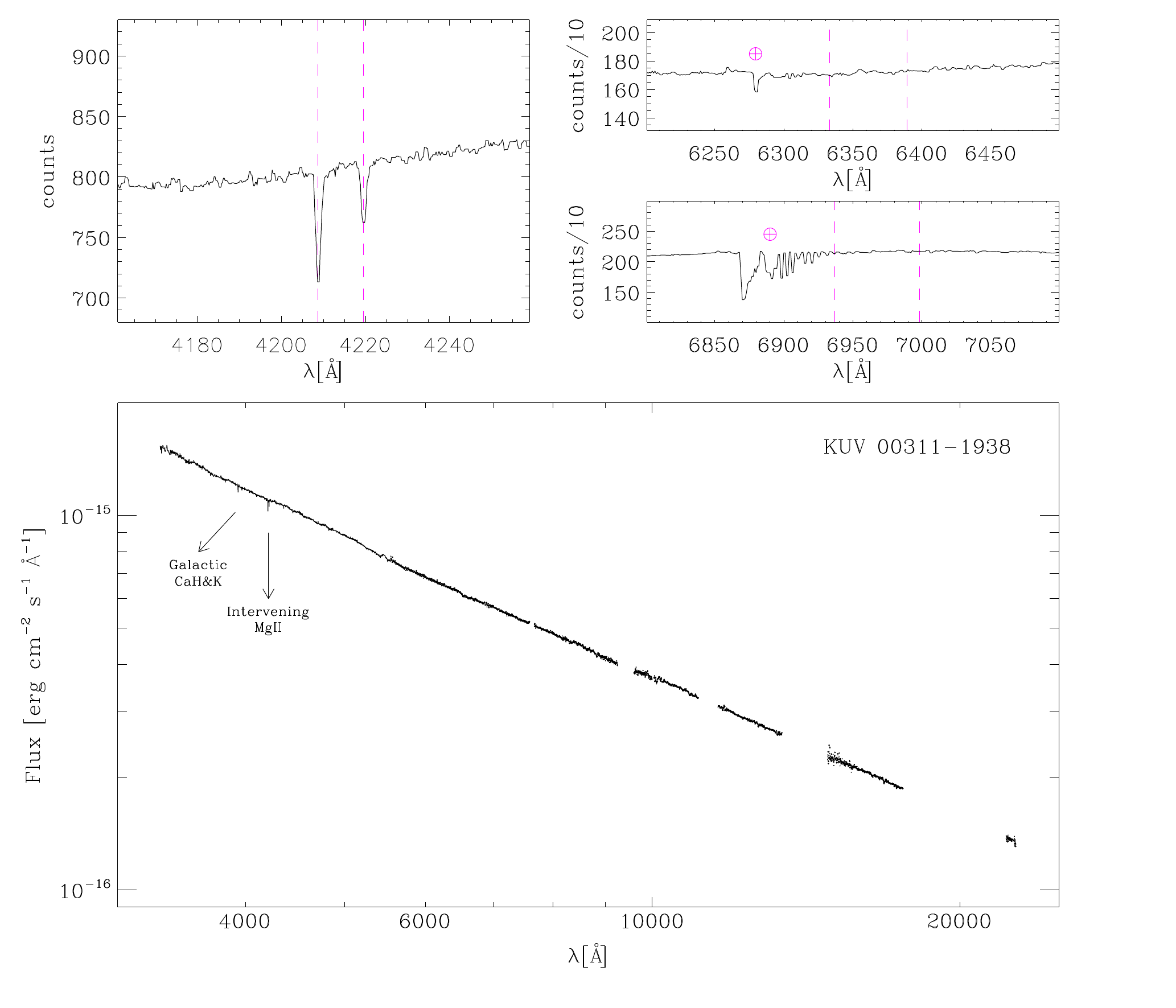}
 \caption{Top: The KUV\,00311--1938 counts spectrum in different regions of interest: where an Mg II doublet, associated to the host galaxy or to an intervening cold system, is identified at $ z = 0.506 $ (left); and where the Ca II H\&K absorption features are expected for the tentative redshifts $z = 0.61$  and $ z = 0.7635$ respectively, as proposed by \citet{Piranomonte_2007:SedentarySurvey} and \citet{Jones_2009:6dFGalaxySurvey} (right). The expected or measured positions for these spectral features are indicated by the dashed vertical lines. Bottom: the UV-to-NIR spectrum of KUV\,00311--1938 obtained with X-shooter after flux calibration and telluric corrections. 
The spectrum is dominated by the non-thermal emission from the nucleus and is well described by a single power law (see Table~\ref{TabResultsLL} for details). Only a single extragalactic spectral feature is identified, a doublet corresponding to the absorption by an Mg II system (see the top-left panel of this figure). The Ca II H\&K doublet associated with our own Galaxy is also identified and indicated in the figure.} 
  \label{FigKUV0033:all}
\end{figure*}

The object KUV\,00311--1938 was initially detected as an ultraviolet-excess during a survey performed with the Kiso Schmidt telescope of the 
Tokyo Astronomical Observatory \citep{Kondo_1984:KUVcatalogue}.
It was later detected as a bright X-ray source and identified as a BL\,Lac by ROSAT \citep{Bauer_2000:ROSAT,Schwope_2000:ROSAT_RBS}.
Subsequently, \citet{Piranomonte_2007:SedentarySurvey} confirmed this identification and proposed a tentative redshift of $z = 0.61$, based on the 
interpretation of very weak features found in a spectrum obtained with the ESO 3.6m telescope. This result is not confirmed in the final 
redshift release of the 6dF Galaxy Survey \citep{Jones_2009:6dFGalaxySurvey}, which reports a tentative redshift (flag Q=2, ``unlikely redshift'') at 0.7635.
The source has recently been detected in $\gamma$-rays, first by \textit{Fermi}/LAT \citep{Abdo_2009:BrightAGN,Ackermann_2011:2LAC}, and later by H.E.S.S. 
\citep{Becherini_2012:Gamma2012}. It is considered as one of the farthest detected BL\,Lac known as VHE $\gamma$-ray emitter.

The observation of KUV\,00311--1938 with X-shooter took place in November 2011 with an exposure time of two hours, resulting in a spectrum with 
a S/N between 25 and 250 depending on the arm (see Table~\ref{tabobs}). 
The near-IR to UV spectrum, shown in the lower panel of Fig.~\ref{FigKUV0033:all}, is reasonably well described by a power law 
with an index (in wavelength) $\alpha = -1.14 \pm 0.14$ and an absolute flux at 10000\,\AA\ of (4.0 $\pm$ 0.2)$\times$10$^{-16}$\,\ergcmsa, 
which indicates that non-thermal emission strongly dominates the host galaxy emission.
The corresponding NIR magnitudes, shown in Table~\ref{magtab}, are close to those from 2MASS \citep{Skrutskie_2006:2MASS}.
 
The search for absorption or emission lines in the whole wavelength range has been done within the 3 cycles of data recording and splitting 
of the data along the slit as indicated in Section~\ref{Observations}. 
The only extragalactic signature found is a doublet of narrow absorption lines (see the upper-left panel of Fig.~\ref{FigKUV0033:all}), which are 
clearly identified in the UVB arm at $\lambda = 4208.7$\,\AA\ and $\lambda = 4219.5$\,\AA.
As shown in \citet{Pita_Gamma2012:XSH}, we identified this as an Mg II doublet at $z = 0.50507$ ($\pm 5 \times 10^{-5}$), a result which was later confirmed by \citet{Shaw_2013:zkeckBLL}. We measured the EWs of the two lines by taking into account the errors in continuum placement as in \citet{SembachAndSavage}. We obtained $0.164 \pm 0.019$\,\AA\ and $0.094 \pm 0.011$\,\AA, respectively.
The ratio of the EWs is thus $1.74 \pm 0.28$ indicating an optically thin cloud.
As there is no indication whether this absorption is produced in the host galaxy or in an intervening cold system, 
this redshift can only be considered as a lower limit of the redshift of the source.
We did not detect any absorption feature of Fe II at the same redshift.
In addition, none of the absorption or emission lines at the tentative redshift proposed by \citet{Piranomonte_2007:SedentarySurvey} ($z = 0.61$) or by \citet{Jones_2009:6dFGalaxySurvey} ($z = 0.7635$) were found. 
This is illustrated in the upper-right panel of Fig.~\ref{FigKUV0033:all}, 
where the regions in which the Ca II H\&K doublet would be located are shown.
As for PKS\,0447--439, we estimated a lower limit on the redshift with the imaging method proposed by \citet{Sbaruf_2005:imared}, using a host galaxy magnitude $M_\mathrm{R} = -22.5$ \citep{Shaw_2013:zbllacs} and $M_\mathrm{R} = -22.9$ \citep{Sbaruf_2005:imared}, thus obtaining, respectively, $z > 0.47$ and $z > 0.5$. 
In addition, from the absence of Lyman $\alpha$ absorption and the S/N near the beginning of the spectrum, we derived an upper limit on the redshift $z_\mathrm{max}$ = 1.54.

\subsection{PKS\,0301--243}
\label{ResultPKS0301}

\begin{figure*}
\centering
 \includegraphics[width=15cm]{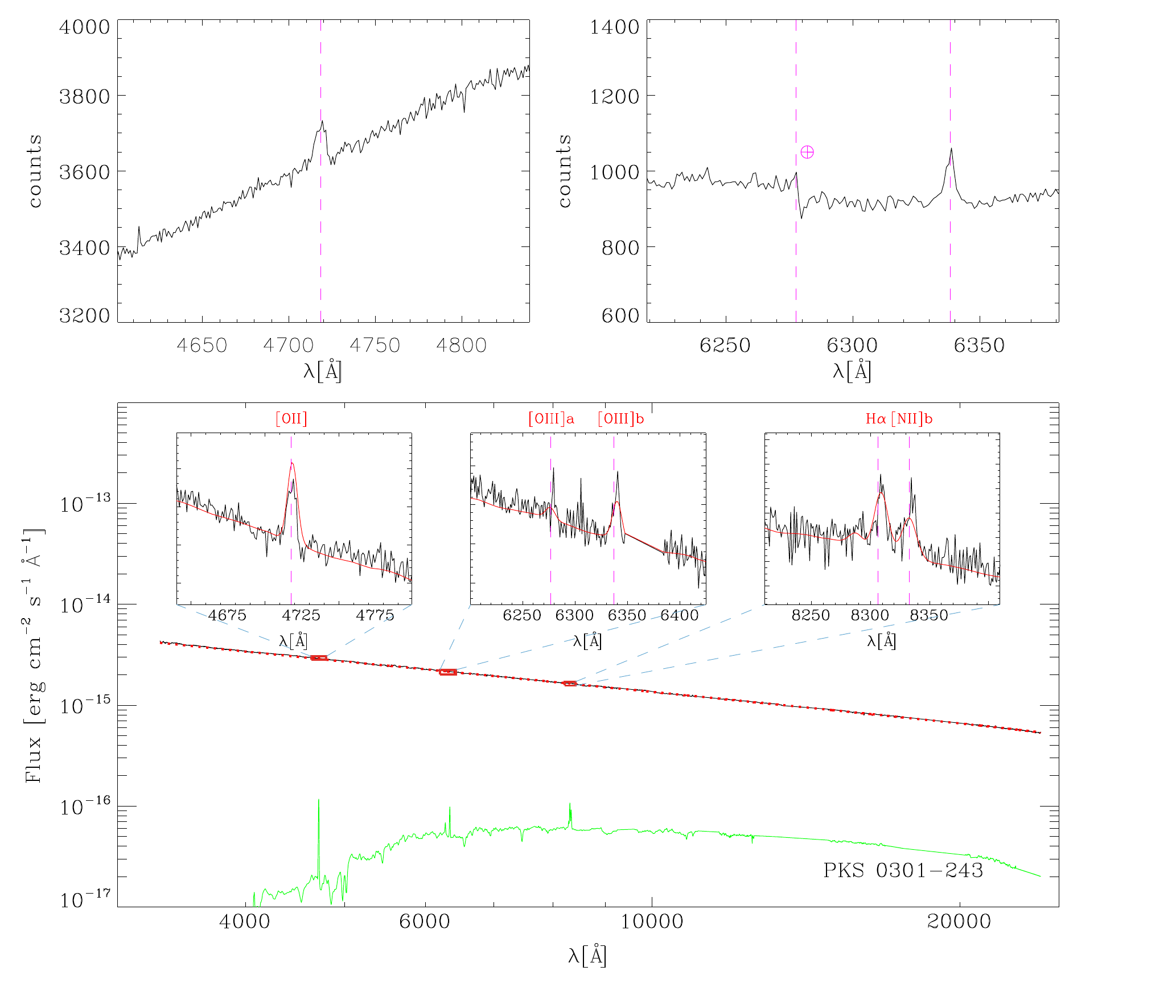}
\caption{Top: Parts of the PKS\,0301--243 counts spectrum showing emission lines in the UVB (left) and VIS (right) arms. For the VIS only the Positive spectrum (see text) is shown, because the line has a better signal-to-noise and because the Central and Negative spectra are affected by bad pixels above 6350\,\AA. As shown by the vertical dashed lines, they correspond to [OII] $\lambda$ 3727\,\AA\ and [OIII] $\lambda$ 5007\,\AA\ emission lines when redshifted to $z = 0.2657$. The [OIII] $\lambda$ 4959\,\AA\ emission line, expected at 6278\,\AA\, is not visible before correction of the telluric absorption. 
Bottom: The UV-to-NIR spectrum of PKS\,0301--243 obtained with X-shooter after flux calibration and telluric corrections (in black) along with the results of our modelling (in red). In the insets, zooms of the emission lines in the X-shooter spectrum along with a local fit (in red). In green the galaxy model, composed by the template of \citet{Mannucci_2001:Template} with added emission lines of fixed width. The template is scaled according to the measured $R$ magnitude \citep{Falomo_2000:BLLac} and to the flux enclosed in the slit, while the emission lines are scaled to the X-shooter flux level (see text for details). Results are given in Tables~\ref{PKS0301lines} and \ref{TabResults}.
}
  \label{FigPKS0301:all}
\end{figure*}

\begin{table*}
\caption{\label{PKS0301lines}PKS\,0301--243 emission lines.}
\centering
\begin{tabular}{cccccc}
\hline\hline
Spectrum & [OII] $\lambda$ 3727  &  [OIII] $\lambda$ 4959 & [OIII] $\lambda$ 5007 & H${\alpha}$ &  [NII] $\lambda$ 6583\\
\hline
& \multicolumn{5}{c}{Line Position (\AA)}  \\
Total &  4718.77 $\pm$ 0.27 & 6277.95 $\pm$ 0.42 &  6336.99 $\pm$ 0.40
& 8307.21   $\pm$     0.47  &  8333.13 $\pm$ 0.60\\
Centre &  4718.44 $\pm$ 0.24 & 6277.65 $\pm$ 0.22 &  6336.95 $\pm$ 0.52
& 8307.31   $\pm$     0.46  & 8333.30 $\pm$ 0.63 \\
Positive &  4718.46 $\pm$ 0.21 & 6277.52 $\pm$ 0.27 &  6337.94 $\pm$ 0.13
& 8307.34   $\pm$     0.24  & n.a. \\
\hline
& \multicolumn{5}{c}{Line EW (\AA)} \\
Total &-0.192 $\pm$  0.020 & -0.094 $\pm$  0.025&  -0.216 $\pm$  0.043 & -0.156 $\pm$
0.034 & -0.17 $\pm$ 0.03\\
Centre & -0.211 $\pm$ 0.020 &-0.064 $\pm$  0.022 &  -0.220 $\pm$  0.043 & -0.191 $\pm$
0.049 & -0.11 $\pm$ 0.03 \\
Positive  & -0.348 $\pm$  0.073 &-0.248 $\pm$  0.074 & -1.022 $\pm$
0.121 & -0.979 $\pm$0.115  & -0.10 $\pm$ 0.06 \\
\hline
\end{tabular}
\tablefoot{Best fit positions and equivalent widths EWs of the emission lines detected in the total, central, and positive spectrum (see text for details).}
\end{table*}

This radio source was discovered during a 1415 MHz continuum survey with the OSU radio telescope \citep{Ehman_1970:OHIO}
and identified as a BL\,Lac by \citet{Impey_1988:Blazar} due to its high optical polarimetric fraction. 
Subsequently, the source has been observed over the entire waveband range from radio to $\gamma$-rays.
It has been discovered as a high energy $\gamma$-ray emitter by \textit{Fermi}/LAT \citep{Abdo_2009:BrightSources,Abdo_2009:BrightAGN} and recently 
at VHE by H.E.S.S. \citep{Wouters_2012:PKS0301,HESS_2013:PKS0301}.

A first estimation of the redshift of the source was proposed by \citet{Pesce_1995:PKS0301} based on the spectroscopic measurement 
of the redshift of two close companions (located at 6\arcsec and 20\arcsec\ from PKS\,0301--243), which are at $z$$\sim$0.263. This suggests that PKS\,0301--243 
could be part of a cluster of galaxies of Abell richness class 0. This result was supported by further observations of PKS\,0301--243
by \citet{Falomo_2000:BLLac}, with the plausible identification of a single weak emission line detected at 6303\,\AA\ with the redshifted 
[O III] 5007\,\AA\ at $z = 0.26$.

The observation of PKS\,0301--243 with X-shooter took place in December 2011 with an exposure time of 4800 seconds resulting in a spectrum with 
a S/N between 35 and 210 depending on the arm  (see Table~\ref{tabobs}). The corresponding NIR magnitudes, shown in Table~\ref{magtab}, are significantly brighter than those from 2MASS \citep{Skrutskie_2006:2MASS}. A first examination of the UVB and VIS spectra allowed the identification of the [OII] $\lambda$ 3727\,\AA\ and [OIII] $\lambda$ 5007\,\AA\ emission lines 
\citep[see the upper panel of Fig.~\protect\ref{FigPKS0301:all} and also][]{Pita_Gamma2012:XSH} and to estimate a redshift of 0.266, which is more precise than 
that by \cite{Falomo_2000:BLLac}.
After telluric corrections, we were also able to detect features corresponding to [OIII] $\lambda$ 4959\,\AA, H$\alpha$, and [NII] $\lambda$ 6583\,\AA\ (see the lower panel of Fig.~\ref{FigPKS0301:all}).
Following \cite{Falomo_2000:BLLac}, we investigated the possibility of an extended origin of these emission lines. To do so, we fitted the median slit 
profile of the UVB and VIS spectra with a Gaussian. Using the results of this fit, we extracted one spectrum (Central) from the region within 
1$\sigma$ (0.6\arcsec in UVB and 0.4\arcsec in VIS) from the peak and two spectra from the two sides at coordinates smaller and bigger than the 
centre (negative and positive). The positive direction is towards ENE. 
Four of the previously listed emission lines were clearly visible in the central spectrum and in the positive spectrum but not in the
negative spectrum. This indicates anisotropic extension of the emitting material. 
The fifth emission line, [NII] $\lambda$ 6548\,\AA, is only visible in the central and in the total spectra.
The EW and best fit position of each emission line were computed fitting the continuum with a cubic spline 
and are shown in Table~\ref{PKS0301lines}.
The emission lines are clearly brighter in the Positive spectrum, indicating extended emission on scales 
larger than 0.6\arcsec\ (1\arcsec\ = 4.09 kpc at $z = 0.266$) in the host galaxy or in its complex environment.
As can be seen, the best fit positions are compatible within the (relatively large) errors. The maximum difference in best fit positions is between 
the [OIII] $\lambda$ 5007\,\AA\ in the central and positive spectrum. Its value in velocity space is 46 $\pm$ 19 km/s, we, therefore, consider that the difference 
in velocity between the emission in the two spectra is less than 57 km/s (3$\sigma$).
To estimate the systemic redshift, we use the [OIII] $\lambda$ 5007\,\AA\ emission line because it is a singlet, in a region free of telluric absorption, and
we take the value from the total spectrum obtaining $z = 0.2657 \pm 8.6 \times 10^{-5}$.

Given the presence of the blazar emission we cannot easily evaluate the flux of the emission lines. To obtain a rough estimate, we modelled the emission of PKS\,0301--243 as a power law plus a galaxy template with added emission lines by following the method described in Section~\ref{FitModel} for the case where the host galaxy continuum is not detected. We obtained a power law wavelength index $\alpha$ = -1.03 $\pm$ 0.02 and an absolute flux at 10000\,\AA\ of (13.7 $\pm$ 0.1)$\times$10$^{-16}$\,\ergcmsa. We estimated the emission of the galaxy inside our slit using the $R$ magnitude and the effective radius from \citet{Falomo_2000:BLLac}. For consistency, the slit losses were estimated  using the enclosed magnitude profile of a de Vaucouleurs profile \citep[e.g.][]{Graham_2005}. We then added Gaussian emission lines with FWHM of 500 km/s to the model to simulate the detected lines, scaling the flux to obtain a reasonable match with the observations. With this rough model, the jet/galaxy ratio \citep{Piranomonte_2007:SedentarySurvey} at 5500\,\AA\ in the rest frame is $\sim$34 in agreement with the almost featureless spectrum we observe. Concerning the emission lines, the ratio ([OII] $\lambda$ 3727\,\AA)/([OIII] $\lambda$ 5007\,\AA) is $\sim$1.5 while the ratio ([NII] $\lambda$ 6583\,\AA)/H$\alpha$ must be slightly smaller than one. Both ratios agree with the hypothesis that the emission lines are generated in a LINER \citep{Ho_2003:Liners}.
 
\subsection{BZB\,J0238--3116}

\begin{figure*}
\centering
 \includegraphics[width=15cm]{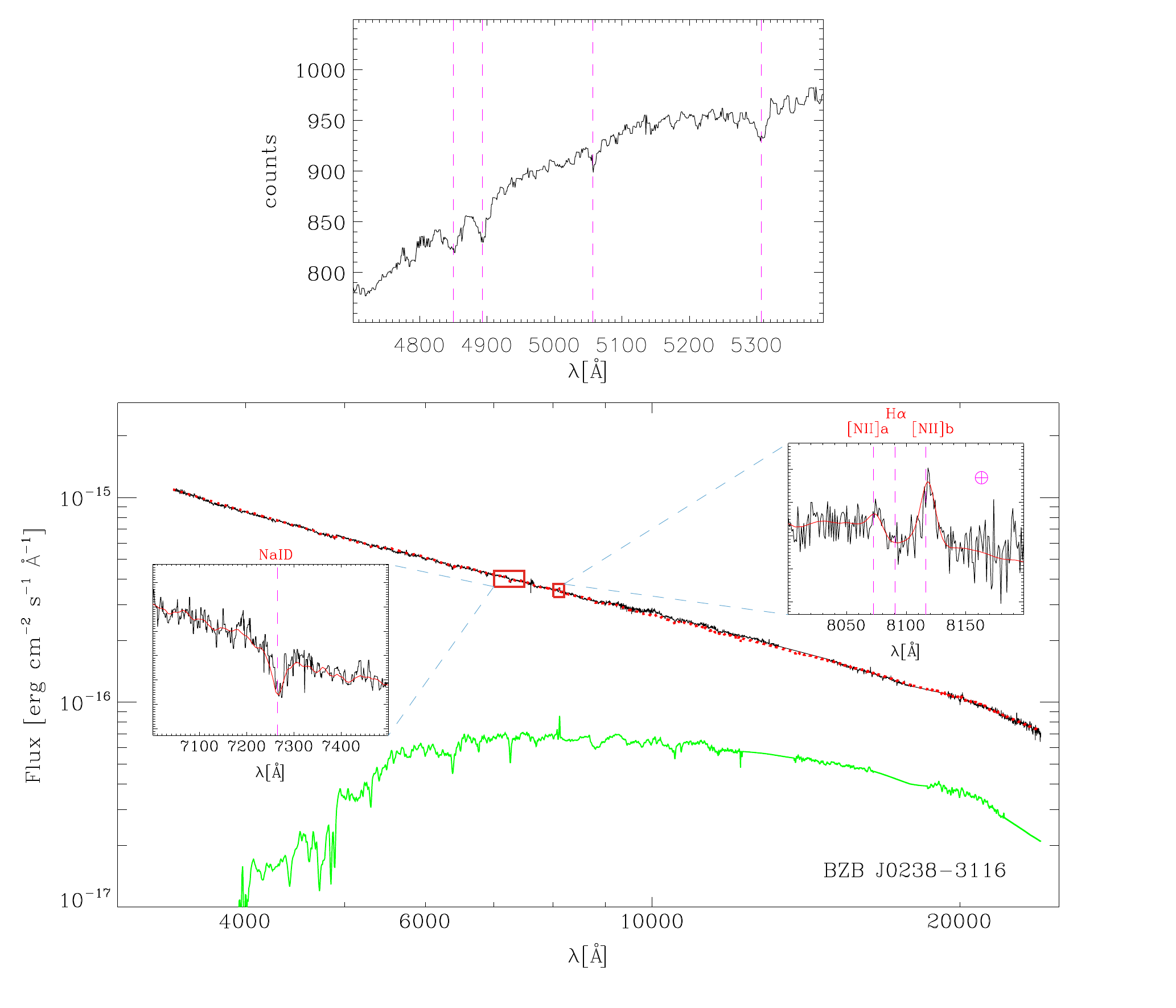}
\caption{
Top: Part of the X-shooter counts spectrum of BZB\,J0238--3116 in the UVB arm after a median filtering with a 5-pixel window. The absorption features are identified as corresponding to the Ca II H\&K, H$\delta$, and Ca I G absorption lines for $z = 0.2328$. The expected positions are indicated by the dashed vertical lines.
Bottom: The UV-to-NIR spectrum of BZB\,J0238--3116 obtained with X-shooter after flux calibration and telluric corrections (in black) along with the results of our modelling (in red). In the insets, zooms are on the absorption and emission lines (Na I D, [NII] $\lambda$ 6583\,\AA\ and H$\alpha$) identified after flux calibration and telluric correction along with a local fit (in red). In green, the galaxy model, as composed by the template of \citet{Mannucci_2001:Template} with added emission lines of fixed width. The galaxy model normalisation and the slope of the model (in red) are the result of a fit, as explained in Section~\ref{FitModel}. Results are given in Table~\ref{TabResults}.}
\label{FigBZBJ0238:all}
\end{figure*}

The X-ray source BZB\,J0238--3116, present in the ROSAT bright source catalogue \citep{Schwope_2000:ROSAT_RBS} and identified as a BL\,Lac in the RBSC-NVSS sample \citep{Bauer_2000:ROSAT}, has been detected as a $\gamma$-ray emitter with a hard spectrum (photon index $\sim$1.85) by \textit{Fermi}/LAT \citep[2FGL J0238.6-3117, see][]{Ackermann_2011:2LAC}.

The observation of BZB\,J0238--3116 with X-shooter took place in December 2011 with an exposure time of
one hour, and the resulting spectrum has a S/N between 25 and 120 depending on the arm (see Table~\ref{tabobs}). The corresponding NIR magnitudes, shown in Table~\ref{magtab}, are close to those from 2MASS \citep{Skrutskie_2006:2MASS}. As already reported in \citet{Pita_Gamma2012:XSH}, a first examination of the spectrum allowed the identification of several features in absorption, namely
Ca II H\&K, H$\delta$, Ca I G, and Na I D (see the upper panel of Fig.~\ref{FigBZBJ0238:all}). Together, these features are consistent with a redshift $z = 0.2328$. 
After flux calibration, we discovered two weak emission features around 8100\,\AA\ that are consistent with the [NII] $\lambda$ 6583\,\AA\ and [NII] $\lambda$ 6548\,\AA\ emission lines, while no emission at the position of H$\alpha$ was detected (see the lower panel of Fig.~\ref{FigBZBJ0238:all}).
Following the procedure described in Section~\ref{ResultPKS0301} for PKS\,0301--243, we measured the equivalent width of these two 
features and obtained EW([NII] $\lambda$ 6583\,\AA) = $-0.71 \pm 0.14$\,\AA\ and  EW([NII] $\lambda$ 6548\,\AA) = $-0.31 \pm 0.13$\,\AA.
Fitting a Gaussian to [NII] $\lambda$ 6583\,\AA, we obtained a redshift $z = 0.2329 \pm 7\times10^{-5}$ consistent with the absorption redshift.

We then modelled the emission of the source following the method described in Section~\ref{FitModel}. The best fit parameters are a power law wavelength index $\alpha$ = -1.57 $\pm$ 0.02, an absolute flux at 10000\,\AA\ of (2.66 $\pm$ 0.01)$\times$10$^{-16}$\,\ergcmsa, and a ratio (power law)/(galaxy) at 5500\,\AA\ in the rest frame $R_{PWL/GAL}$ = 6.0 $\pm$ 0.4 (see Table~\ref{TabResults}). We then added two Gaussian lines with 500 km/s FWHM to the best  fit model at the position of the [NII] $\lambda$ 6583\,\AA\ and [NII] $\lambda$ 6548\,\AA\ emission lines, and we scaled their fluxes to match the observations maintaining a ratio of 3:1 between their fluxes. The result of this modelling is shown in the upper panel of Fig.~\ref{FigBZBJ0238:all}. The implied absolute magnitude of the galaxy is $M_\mathrm{R} = -22.5 \pm 0.1$, which is compatible with the value reported in \citet{Shaw_2013:zbllacs}.

The redshift of BZB\,J0238--3116, which was unknown at the moment of the observations reported here, has been confirmed independently in two recent papers: \citet{Landoni_2013:zbllacs} from observations performed with the FORS2 spectrograph at the VLT, and \citet{Shaw_2013:zbllacs} from observations performed with the WMKO telescope at W. M. Keck Observatory. Both papers report a value of $z = 0.232$, which is in good agreement with $z = 0.2329$ reported in this paper.

\subsection{BZB\,J0543--5532}

\begin{figure*}
\centering
 \includegraphics[width=15cm]{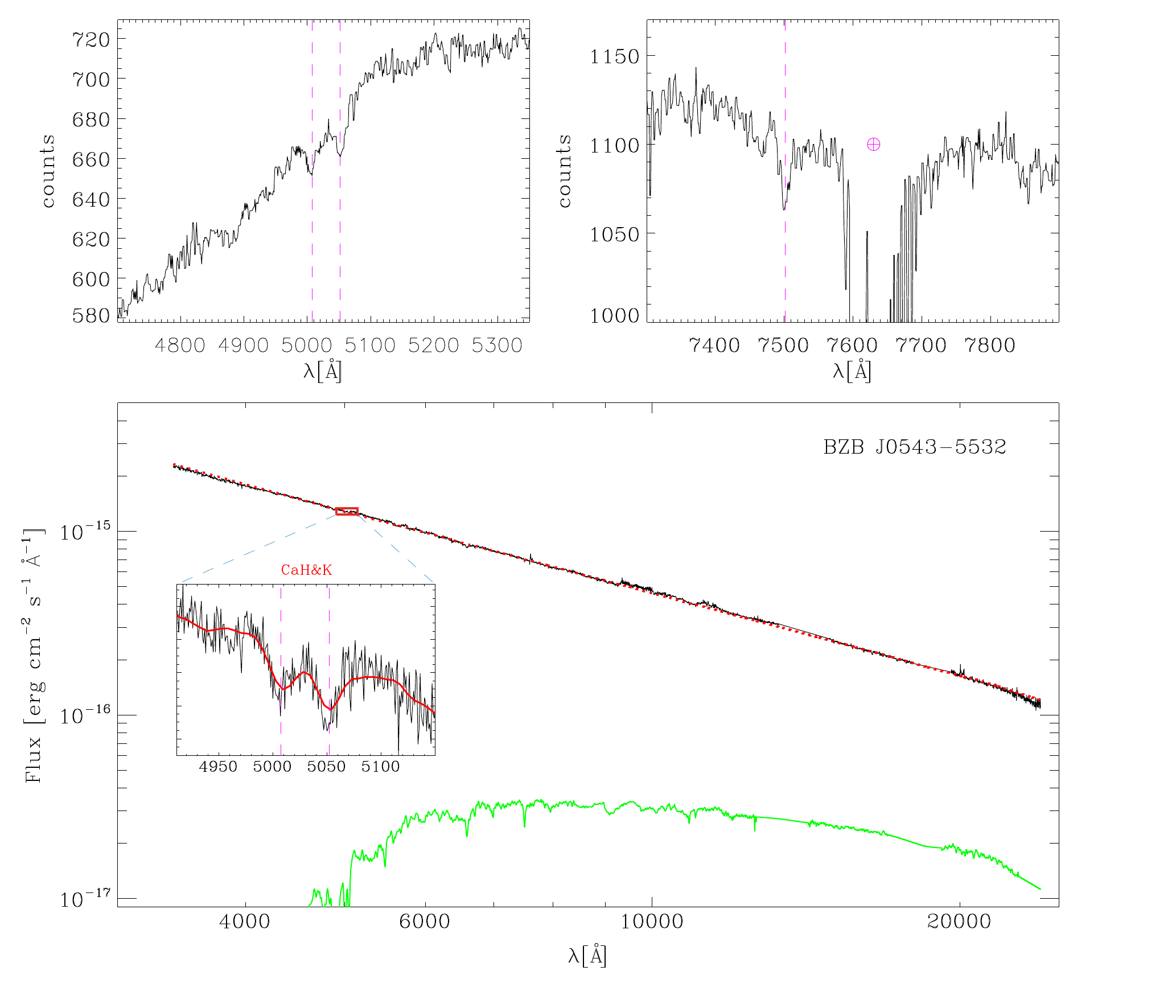}
\caption{Top: Parts of the X-shooter counts spectrum of BZB\,J0543--5532 both in the UVB and VIS arms, after a median filtering with a 3-pixel window. The absorption features are identified as corresponding to the Ca II H\&K (left) and Na I D (right) for $z = 0.2730$. The expected positions are indicated by the dashed vertical lines. 
Bottom: The UV-to-NIR spectrum of BZB\,J0543--5532 (in black) along with the result of our modelling (in red), as based on a fit of a galaxy template (in green) and a power law (see Section~\ref{FitModel} for details and Table~\ref{TabResults} for results). In the inset, a zoom of the absorption lines corresponding to Ca II H\&K doublet (already identified in the counts spectrum and shown in the upper panel) and a local fit (in red) are shown.}
  \label{FigBZBJ0543:all}
\end{figure*}

The X-ray bright source BZB\,J0543--5532, detected and optically identified as a BL\,Lac by ROSAT \citep{Fischer_1999:ROSAT}, was later classified as HBL by \citet{Anderson_2009:RadioBL} based on radio observations. The source has been discovered in $\gamma$-rays with \textit{Fermi}/LAT \citep{Abdo_2010:1FGL} and shows a hard spectrum \citep[photon index $\sim$1.74, see][]{Ackermann_2011:2LAC}. 

The redshift of this source is unknown. A photometric lower limit $z > 0.27$ ($z > 0.38$) is derived by \citet{Shaw_2013:zbllacs}, assuming $M_\mathrm{R} = -22.5$ ($M_\mathrm{R} = -22.9$ respectively) for the host galaxy, and a photometric upper limit $z < 1.08$ is proposed by \citet{Rau_2012:Zphot} from the fit of a 13-bands UV to NIR photometry.

The observation of BZB\,J0543--5532 with X-shooter took place in December 2011 with an exposure time of 1.5 h, and the resulting spectrum has an average S/N between 20 and 110 depending on the arm (see Table~\ref{tabobs}). The corresponding NIR magnitudes, shown in Table~\ref{magtab}, are slightly fainter than those from 2MASS \citep{Skrutskie_2006:2MASS}. A first examination \citep[see][]{Pita_Gamma2012:XSH} of the counts spectrum allowed us to detect the Ca II H\&K doublet and the Na I D absorption lines at a redshift $z = 0.2730$, as shown in the upper panel of Fig.~\ref{FigBZBJ0543:all}. We could detect no emission lines. 

We then modelled the emission of the source following the method described in Section~\ref{FitModel}. In this case, we obtained a power law wavelength index $\alpha$ = $-1.56 \pm 0.10$, an absolute flux at 10000\,\AA\ of (4.60 $\pm$ 0.01)$\times$10$^{-16}$\,\ergcmsa, and a ratio (power law)/(galaxy) at 5500\,\AA\ in the rest frame $R_{PWL/GAL}$ = 20.0 $\pm$ 0.6. The derived absolute magnitude of the galaxy is $M_\mathrm{R} = -21.4 \pm$ 0.6.

\subsection{BZB\,J0505+0415}

\begin{figure*}
\centering
 \includegraphics[width=15cm]{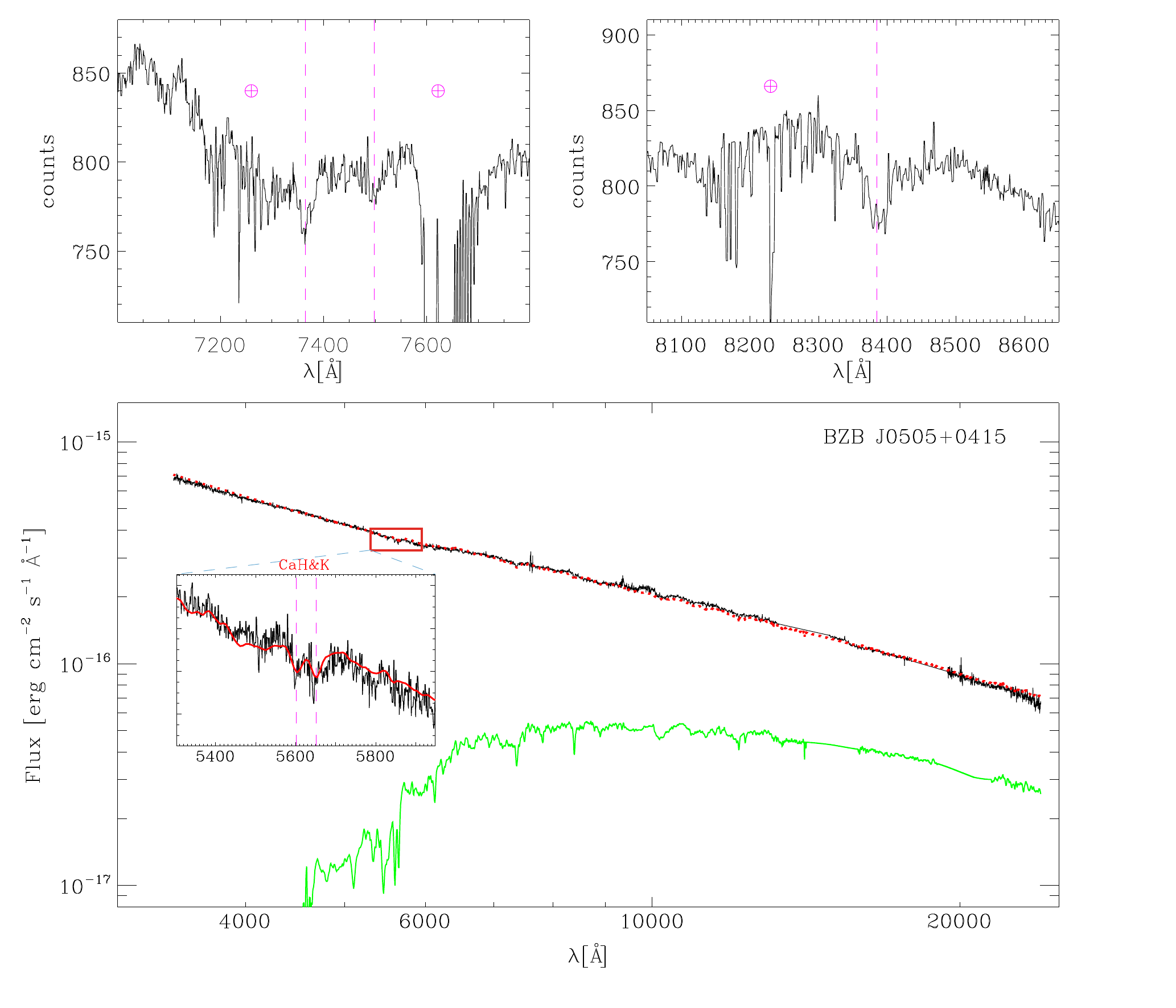}
\caption{Top: Parts of the X-shooter counts spectrum of BZB\,J0505+0415 in the VIS arm after a median filtering with a 3-pixel window. The absorption features are identified as corresponding to the Mg and CaI+FeI (left) and Na I D (right) lines for $z = 0.424$. The expected positions are indicated by the dashed vertical lines. 
Bottom: the UV-to-NIR spectrum of BZB\,J0505+0415 (in black) along with the result of our modelling (in red), based on a fit of a galaxy template (in green) and a power law (see Section~\ref{FitModel} for details, and Table~\ref{TabResults} for results). In the inset, a zoom of the absorption lines corresponding to Ca II H\&K doublet at $z = 0.424$, identified after flux calibration and telluric correction, is shown along with a local fit (in red).}
  \label{FigBZBJ0505:all}
\end{figure*}

The object BZB\,J0505+0415 was first discovered as a radio source within the Parkes-MIT-NRAO survey \citep{Griffith_1995:PMN} and later in X-rays during the ROSAT survey \citep{Brinkmann_1997:ROSAT}. It has been spectroscopically identified as a BL\,Lac by \citet{Laurent-Muehleisen_1998:ROSAT}. Recently, the source has been detected in $\gamma$-rays with \textit{Fermi}/LAT \citep{Nolan_2012:2FGL}.

The redshift of the source is unknown. A first tentative redshift is proposed at $z = 0.0272$ by \citet{Bauer_2000:ROSAT}, but is considered as uncertain: the spectrum is of poor quality, and such a small value would imply a very faint and small host galaxy \citep{Nilsson_2003:BLLacs}. Recently, a photometric  lower limit $z > 0.32$ ($z > 0.34$) is derived by \citet{Shaw_2013:zbllacs} assuming $M_\mathrm{R} = -22.5$ ($M_\mathrm{R} = -22.9$ respectively) for the host galaxy and an upper limit $z < 1.26$ has been proposed by \citet{Rau_2012:Zphot} from the fit of a 13-bands UV to NIR photometry. The apparent magnitude of the host galaxy has been measured as $m_\mathrm{R} = 18.05 \pm 0.04$ with a fit of the profile using a de Vaucouleurs profile \citep{Nilsson_2003:BLLacs}.

The observation of BZB\,J0505+0415 with X-shooter took place in December 2011 with an exposure time of two hours and the resulting spectrum has an average S/N between 20 and 90 depending on the arm (see Table~\ref{tabobs}).  The corresponding NIR magnitudes, as shown in Table~\ref{magtab}, are close to those from 2MASS \citep{Skrutskie_2006:2MASS}. As already reported in \citet{Pita_Gamma2012:XSH}, a first examination of the spectrum allowed the identification of several features in absorption, namely Mgb, CaI+FeI, and Na I D at redshift $z = 0.424$ (see the upper panel of Fig.~\ref{FigBZBJ0505:all}). After flux calibration and merging of the spectra, we were also able to identify the Ca II H\&K doublet, which falls at the junction between the UVB and VIS arm. We could not detect any emission line. However, we remark that the complex with the H$\alpha$ emission line and the [NII] doublet falls in a region of high telluric absorption.

The spectrum of the source (see the lower panel of Fig.~\ref{FigBZBJ0505:all}) is then modelled following the method described in Section~\ref{FitModel}. In this case, we obtained a power law wavelength index $\alpha$ = -1.42 $\pm$ 0.10, an absolute flux at 10000\,\AA\ of (2.06 $\pm$ 0.02)$\times$10$^{-16}$\,\ergcmsa, and a ratio (power law)/(galaxy) at 5500\,\AA\ in the rest frame $R_{PWL/GAL}$ = 4.5 $\pm$ 0.8. We applied slit-loss corrections using the effective radius of \citet{Nilsson_2003:BLLacs} obtaining an apparent magnitude for the host galaxy $m_\mathrm{R} = 18.6 \pm 0.2$, which is compatible with the photometric value given above. The final absolute magnitude of the galaxy is $M_\mathrm{R} = -24.2 \pm 0.2$, indicating a very luminous galaxy.

\subsection{BZB\,J0816--1311}

\begin{figure*}
\centering
 \includegraphics[width=15cm]{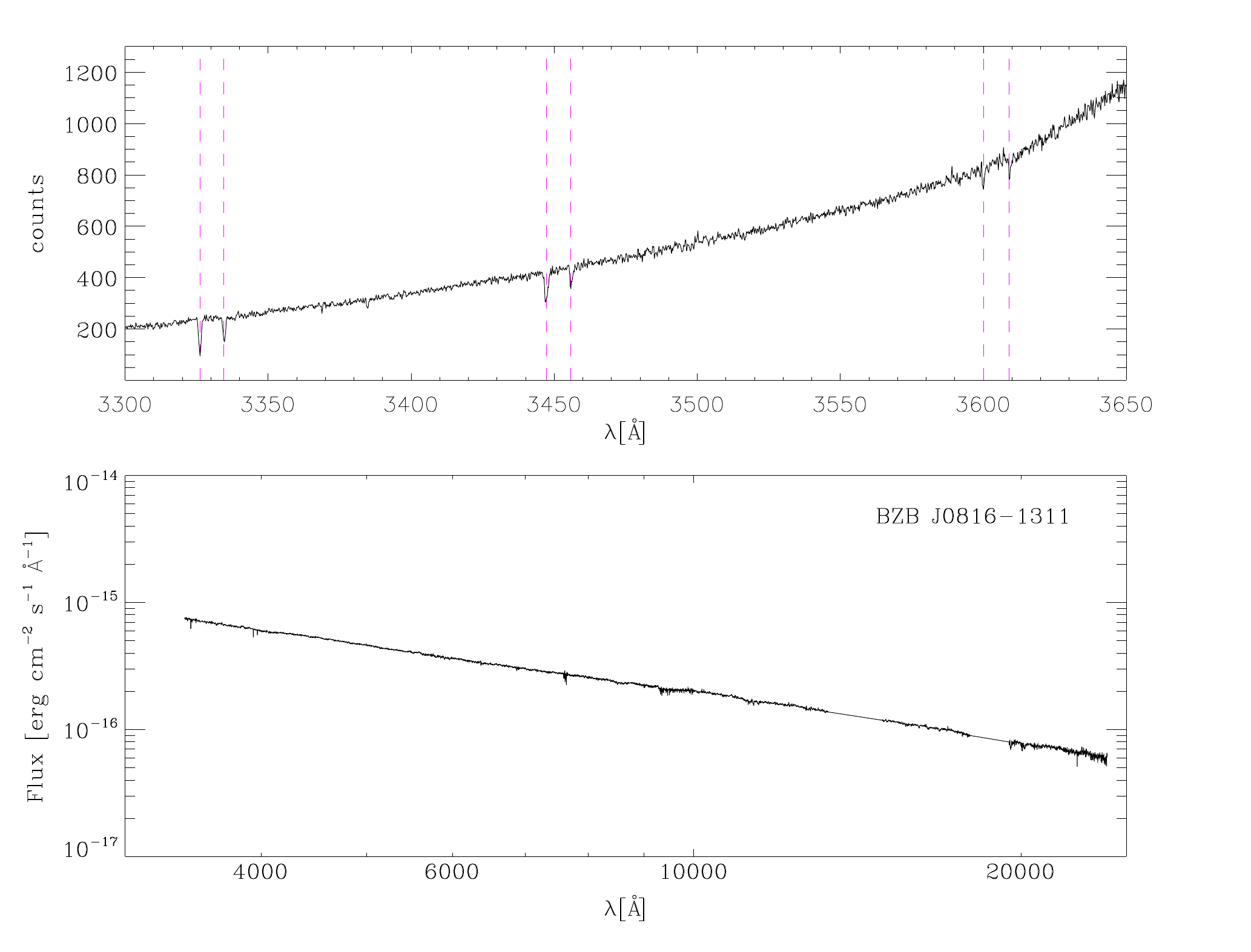}
\caption{Top: Part of the X-shooter counts spectrum of BZB\,J0816--1311 in the UVB arm before flux calibration and telluric corrections. The absorption features are identified as corresponding to the Mg II doublet in three intervening systems, respectively at $z = 0.1902$, $z = 0.2336$, and $z = 0.2882$. The corresponding positions are indicated by the dashed vertical lines. 
Bottom: The UV-to-NIR spectrum of BZB\,J0816--1311 obtained with X-shooter after flux calibration and telluric corrections. The spectrum is dominated by the non-thermal emission from the nucleus and is well described by a power law (see Table~\ref{TabResultsLL} for details). No other extragalactic spectral features are identified.}
  \label{FigBZBJ0816:all}
\end{figure*}

\begin{table*}
\caption{\label{MgII:BZBJ0816}Mg II absorbing systems in the line of sight of BZB\,J0816--1311.}
\centering
\begin{tabular}{ccccc}
\hline\hline
Redshift & EW$_{1}$ (\AA) & EW$_{2}$ (\AA) & Ratio & S \\  
\hline
0.1902  & 0.64 $\pm$ 0.06 & 0.38 $\pm$ 0.05 &  1.68 $\pm$ 0.09 & 11.8 $\sigma$\\
0.2336  & 0.39 $\pm$ 0.06 & 0.21 $\pm$ 0.06 & 1.85 $\pm$ 0.09 & 6.7 $\sigma$\\
0.2882  & 0.07 $\pm$ 0.02 & 0.08 $\pm$ 0.03 & 1 - 0.02& 4.8 $\sigma$ \\
\hline
\end{tabular}
\tablefoot{Redshift, equivalent widths (EW$_{1}$ and EW$_{2}$) of both components of the Mg II absorbing systems, their ratio, and the summed signal-to-noise (S) of their detection.}
\end{table*}

Discovered by ROSAT in X-rays, BZB\,J0816--1311 has been identified as a BL\,Lac object by \citet{Motch_1998:IdROSAT}. It has been discovered in $\gamma$-rays with \textit{Fermi}/LAT \citep{Abdo_2010:1FGL} and shows a hard spectrum in the HE band with a photon index $\sim$1.80 \citep{Ackermann_2011:2LAC}. In the final redshift release of the 6dF galaxy survey \citep{Jones_2009:6dFGalaxySurvey}, a redshift at $z = 0.046$ is proposed for this source, with a quality status $Q=3$, which is described as 'probable'. Recently \citet{Shaw_2013:zbllacs} derived a  photometric lower limit on the redshift  $z > 0.28$ ($z > 0.34$) by assuming $M_\mathrm{R} = -22.5$ ($M_\mathrm{R} = -22.9$ respectively) for the host galaxy and an upper limit $z < 1.66$ from the absence of Lyman $\alpha$ absorption in their spectra.

The observation of BZB\,J0816--1311 with X-shooter took place in November 2011 with an exposure time of 6000 seconds, which results  
in a spectrum with a S/N between 20 and 110 depending on the arm (see Table~\ref{tabobs}). 
As shown in Table~\ref{magtab}, the corresponding NIR magnitudes are slightly fainter than those from 2MASS \citep{Skrutskie_2006:2MASS}.
The NIR to UV spectrum, as shown in Fig.~\ref{FigBZBJ0816:all}, is reasonably well described by a power law 
with an index of -1.24 $\pm$ 0.15 and an absolute flux at 10000\,\AA\ of (2.0 $\pm$ 0.1)$\times$10$^{-16}$\,\ergcmsa, 
which indicates that non-thermal emission dominates the host galaxy emission.
The search for absorption or emission lines in the whole wavelength range has been done within the three cycles of data recording 
and splitting of the data along the slit as indicated in Section~\ref{Observations}. 
As a result and as shown in the upper panel of Fig.~\ref{FigBZBJ0816:all}, only three doublets of narrow absorption lines have 
been identified in the UVB arm.
The three doublets are identified as the absorption by Mg II; as for each doublet the ratio between the two lines positions 
is 1.0025. The corresponding redshifts are: 0.1902, 0.2336 and 0.2882. 
The first two systems are clearly detected and are optically thin, the third one is fainter and appears to be
saturated. The characteristics of the three Mg II doublets are given in Table~\ref{MgII:BZBJ0816}. 
No other absorption features were detected at these redshifts, in particular corresponding to Fe II.

As there is no indication if the absorption at $z = 0.2882$ is produced in the host galaxy or in an intervening cold system, this redshift can only be considered as a lower limit of the redshift of the source ($z \ge$ 0.2882). From the absence of Lyman $\alpha$ absorption and considering the S/N near the beginning of the spectrum, we derive an upper limit to the redshift $z < 1.56$.

\subsection{RBS\,334}

\begin{figure*}
\centering
 \includegraphics[width=15cm]{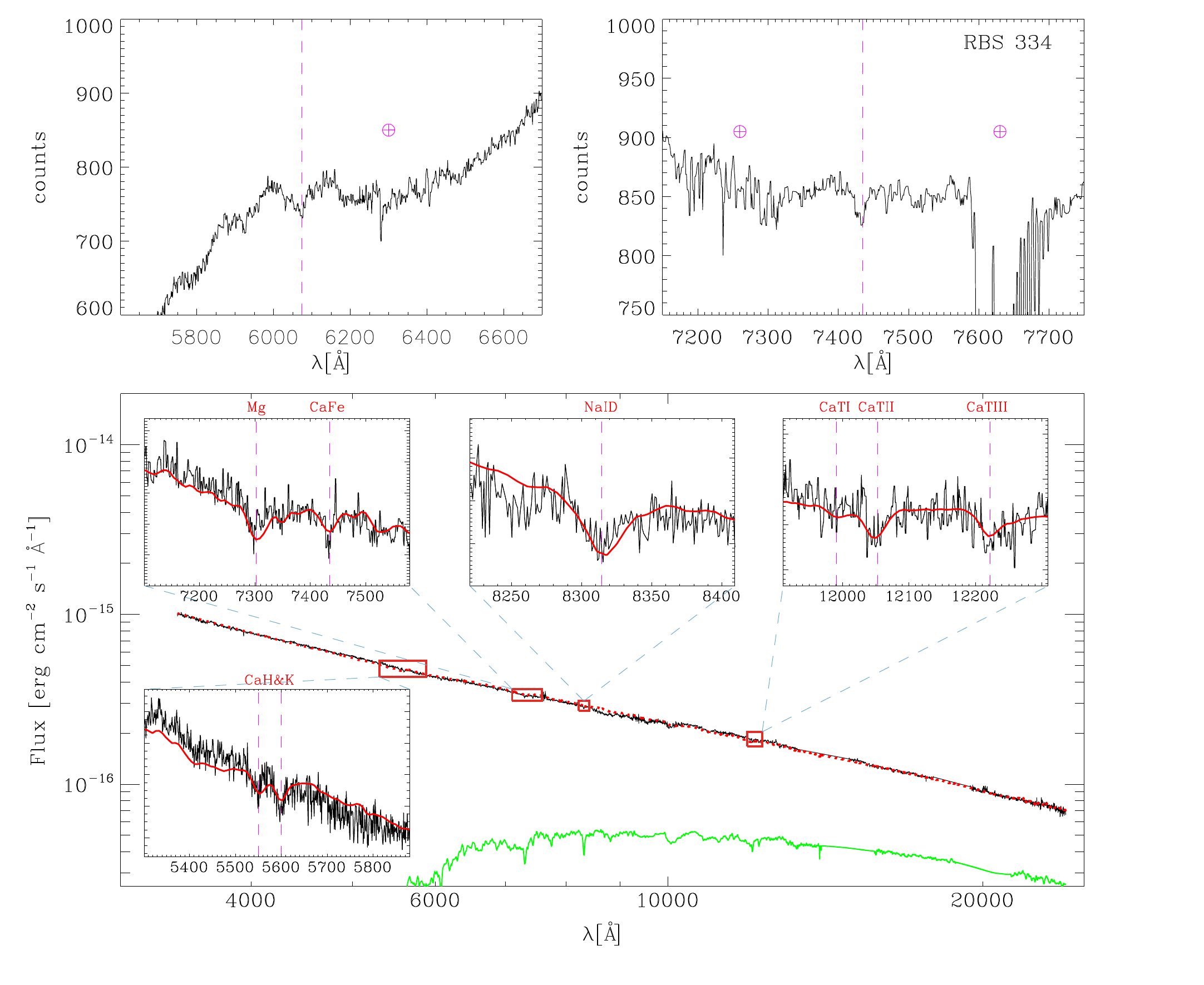}
\caption{Top: Part of the X-shooter counts spectrum of RBS\,334 in the VIS arm after a median filtering with a 3-pixel window. The absorption features are identified as corresponding to the Ca I G (left) and CaI+FeI (right) lines for $z = 0.411$. The expected positions are indicated by the dashed vertical lines. 
Bottom: The UV-to-NIR spectrum of RBS\,334 (in black) along with the result of our modelling (in red) based on a fit of a galaxy template (in green) and a power law (see Section~\ref{FitModel} for details and Table~\ref{TabResults} for results). In the insets, zooms in regions where different absorption features were identified at $z = 0.411$ after flux calibration and telluric correction, which corresponds to Ca II H\&K, Mg, CaI+FeI, Na I D, and the two stronger absorption lines of the Calcium triplet.}
 \label{FigRBS334:all}
\end{figure*}

The X-ray source RBS\,334, present in the ROSAT Bright Source Catalogue \citep{Schwope_2000:ROSAT_RBS}, was identified as a BL\,Lac object in the RBSC-NVSS sample \citep{Bauer_2000:ROSAT}.
It has been discovered in $\gamma$-rays with \textit{Fermi}/LAT \citep{Abdo_2010:1FGL} and shows a very hard spectrum in the HE band with a photon index $\sim$1.56 \citep[see][]{Ackermann_2011:2LAC}. The redshift of the source is unknown.

This source is the only one in our sample which was observed in November 2010. The exposure time was 4800 seconds for an average S/N between 20 and 120 depending on the arm (see Table~\ref{tabobs}). The corresponding NIR magnitudes, as shown in Table~\ref{magtab}, are close to those from 2MASS \citep{Skrutskie_2006:2MASS}.
In \citet{Pita_Gamma2012:XSH} we reported a possible redshift, $z = 0.411$, on the basis of the detection of the Ca II H\&K feature split between the UVB and VIS arms. As shown in Fig.~\ref{FigRBS334:all} after flux calibration and merging of the arms, we are able to confirm this detection and, therefore, the proposed redshift. Moreover, after correction of the telluric absorption, we detected features corresponding to the presence of Ca I G, Mg, CaI+FeI, and Na I D in the VIS spectrum, and the two stronger absorption lines of the Calcium triplet in the NIR spectrum.
No emission lines could be detected. We then modelled the emission of the source as described in Section~\ref{FitModel}, obtaining the best fit with a power law wavelength index $\alpha$ = $-1.66 \pm$ 0.15, an absolute flux at 10000\,\AA\ of (2.26 $\pm$ 0.02)$\times$10$^{-16}$\,\ergcmsa, and a ratio (power law)/(galaxy) at 5500\,\AA\ in the rest frame, $R_{PWL/GAL}$ = 5.0 $\pm$ 1.6. The derived absolute magnitude of the galaxy is $M_\mathrm{R} = -24.0 \pm 0.3$.

\section{Discussion}
\label{Discussion}

\begin{table*}
\caption{\label{TabResLines}List of main features detected for each source}
\centering
\begin{tabular}{llll}
\hline\hline
Source                 & Redshift  & Absorption lines & Emission lines \\
\hline
BZB\,J0238--3116   & 0.2329  &  Ca II H\&K, H${\delta}$, Ca I G, Na I D &  [NII] $\lambda$ 6548\,\AA,  [NII] $\lambda$ 6583\,\AA\\
BZB\,J0543--5532   & 0.273   &  Ca II H\&K, Na I D\\
BZB\,J0505+0415    & 0.424   &  Ca II H\&K, Mgb, CaI+FeI, Na I D\\
RBS\,334           & 0.411   &  Ca II H\&K, Ca I G, CaI+FeI, Na I D, Ca triplet\\
PKS\,0301--243     & 0.2657  &  & [OII] $\lambda$ 3727\,\AA, [OIII] $\lambda$ 4959\,\AA, [OIII] $\lambda$ 5007\,\AA\\
                   &         &  & [NII] $\lambda$ 6548\,\AA, H${\alpha}$, [NII] $\lambda$ 6583\,\AA\\
\hline
KUV\,00311--1938   & 0.506$\le$z$\le$1.54  & Mg II $\lambda$ 2796\,\AA, 2803\,\AA \\
BZB\,J0816--1311   & 0.288$\le$z$\le$1.56  & Three Mg II $\lambda$ 2796\,\AA, 2803\,\AA\ at different redshifts\\
\hline
\end{tabular}
\tablefoot{The columns contain : (1) the name of the source, (2) the redshift, (3 and 4) the name of the absorption and emission features detected. See Section~\ref{Results} for details.}
\end{table*}

\begin{table*}
\caption{\label{TabResults}Main results of our analysis for sources with measured redshift.}
\centering
\begin{tabular}{cccccccc}
\hline\hline
Source                 & Redshift  & Detected & Flux (\ergcmsa) & Power law index & Flux ratio & $m_\mathrm{R}$(galaxy) &  $M_\mathrm{R}$ (galaxy) \\
                       &           & line type  & 10000\,\AA  &            & 5500\,\AA &                        &        \\
\hline
BZB\,J0238--3116   & 0.2329   &  Abs, Em & 2.66($\pm$0.01) $\times$ 10$^{-16}$  & -1.57 $\pm$ 0.02  & 6.0 $\pm$ 0.4   & 18.5 $\pm$ 0.1 & -22.5 $\pm$ 0.1\\
BZB\,J0543--5532   & 0.273    &  Abs   & 4.60($\pm$0.01) $\times$ 10$^{-16}$  & -1.56 $\pm$ 0.10  & 25.0 $\pm$ 10.0 & 20.0 $\pm$ 0.6 & -21.4 $\pm$ 0.6 \\
BZB\,J0505+0415    & 0.424    &  Abs   & 2.06($\pm$0.02) $\times$ 10$^{-16}$  & -1.42 $\pm$ 0.10  & 4.5 $\pm$ 0.8   & 18.6 $\pm$ 0.2 & -24.2 $\pm$ 0.2  \\
RBS\,334           & 0.411    &  Abs   & 2.26($\pm$0.02) $\times$ 10$^{-16}$  & -1.66 $\pm$ 0.15  & 5.0 $\pm$ 1.6   & 18.7 $\pm$ 0.3 & -24.0 $\pm$ 0.3  \\
PKS\,0301--243     & 0.2657   &  Em   & 13.7($\pm$0.1)  $\times$ 10$^{-16}$  & -1.03 $\pm$ 0.02  & {$\sim$}34      & 18.7 (-0.4)    & -23.4  \\
\hline
\end{tabular}
\tablefoot{The columns contain from left to right: (1) source name, (2) redshift, (3) detected line type, absorption (Abs) or emission (Em), (4) flux at 10000\,\AA, (5) UVB--NIR power law slope, (6) jet/galaxy ratio at 5500\,\AA\ in the rest frame, (7) apparent magnitude of the host galaxy before slit loss correction (for PKS\,0301-243, we quote the magnitude measured by \citet{Falomo_2000:BLLac} and the upper limit derived following the discussion in Section~\ref{FitModel}), and (8) absolute magnitude of the host galaxy corrected by slit losses considering an effective radius of 10 kpc for all sources, except BZB\,J0505+0415, for which we used the effective radius from \citet{Nilsson_2003:BLLacs}, and PKS\,0301-243, for which we quote the magnitude measured by \citet{Falomo_2000:BLLac}.}
\end{table*}

\begin{table*}
\caption{\label{TabResultsLL}Main results of our analysis for sources without measured redshift.} 
\centering
\begin{tabular}{cccccccc}
\hline\hline
Source                 & Redshift                & Detected &   Redshift     & Flux (\ergcmsa)       & Power law index & Flux ratio  & m$_R$(galaxy) \\
                       &   spectroscopic         & line type  & imaging        & 10000\,\AA  &                            & 5500\,\AA &                          \\
\hline
KUV\,00311--1938  & 0.506 $\le$ z $\le$ 1.54   & Int & $\ge$ 0.47 &  4.0($\pm$0.2) $\times$ 10$^{-16}$  & -1.14 $\pm$ 0.14  & $\ge$  18 & $\le$ 19.7  \\
BZB\,J0816--1311   &  0.288 $\le$ z $\le$ 1.56  & Int & $\ge$ 0.33   & 2.0($\pm$0.1) $\times$ 10$^{-16}$ & -1.24 $\pm$ 0.15  & $\ge$ 10  & $\le$ 19.5  \\
PKS\,0447--439      &  z $\le$ 1.51 & None & $\ge$ 0.175  & 38.7($\pm$1.6) $\times$ 10$^{-16}$  & -1.32 $\pm$ 0.09 & $\ge$ 20 &  $\le$ 16.7\\
\hline
\end{tabular}
\tablefoot{The columns contain from left to right: (1) source name, (2) spectroscopic limits on redshifts (lower limits based on intervening systems and upper-limits on the non detection of Ly$\alpha$), (3) detected line type, ``Int'' for intervening system, (4) limits on redshifts based on the imaging method proposed by \citet{Sbaruf_2005:imared} (for $M_\mathrm{R}$ = $-22.5$), (5) the flux at 10000\,\AA, (6) UVB--NIR power law slope, (7) limits on the jet/galaxy ratio at 5500\,\AA\ in the determined rest frame that assumes a redshift at 0.506, 0.288, and 0.175, respectively, for KUV\,00311--1938, BZB\,J0816--1311, and PKS\,0447--439, and (8) upper limit to the apparent magnitude of the host galaxy.}
\end{table*}

We have observed a sample of eight BL\,Lac objects with unknown or uncertain redshift, which were previously believed to lie at $z \ge 0.2$, with the new X-shooter spectrograph to determine or better constrain the redshift by detecting absorption or emission features of the host galaxy or absorbing systems in the line of sight (see Table~\ref{TabResLines} for the list of the features we detected for each source). 
With this new instrument, we hoped to be substantially more efficient than previous observations thanks to the higher resolution in a wide waveband with respect to previous observations. In this work, the importance of a higher resolution has been clearly demonstrated by the invalidation of the proposed redshifts for PKS\,0447--439 and BZB\,J0816--1311. The benefit of a wide waveband is illustrated by the detection in this work of several features in the UVB and VIS arms, so allowing the unambiguous determination of several redshifts. This is, for example, the case of PKS\,0301--243, for which we could detect five emission features compared to one detected by \citet{Falomo_2000:BLLac}. The very good S/N allowed us to measure the 5500 \AA\ rest frame jet/galaxy ratio for four sources, up to a value as high as 25. However, this value should not be considered as a general limit, as it depends on several parameters, such as the the S/N, the redshift, and the slope of the power law. For comparison, \citet{Piranomonte_2007:SedentarySurvey} on the basis of simulations, proposed that the BL\,Lac spectra become featureless when this ratio exceeds 10.

\subsection{Value of the NIR arm}
The contribution of the NIR arm to the detection of spectral features has been quite low with only the detection of the Calcium triplet in the case of RBS\,334. There are several reasons for this. On the one hand, the NIR band is affected by the presence of several strong atmospheric absorption bands and emission lines. Moreover, the technological level of NIR detectors is not as good as the one of CCDs used in the UVB and VIS arms, as demonstrated by the presence of many more cosmetic defects (such as hot and cold pixels). On the other hand, concerning the characteristics of the source, the lower intrinsic flux implies a lower signal-to-noise ratio in this part of the spectrum. More importantly, as the redshifts of BL\,Lac objects detected by \textit{Fermi}/LAT are mostly smaller than 1 \citep{Ackermann_2011:2LAC}, the absorption features which can be found in the NIR band are much weaker than those present in the UVB and VIS bands.  

Concerning emission lines, the NIR arm may allow the detection of the H$\alpha$ for redshifts greater than 0.5. Indeed, if we consider our model of PKS\,0301-243, which includes the fit of H$\alpha$, and we displace the source at redshifts greater than 0.5, we find that, accounting for the evolution of the S/N with the wavelength, this emission line remains detectable in the NIR arm up to redshifts z$\sim$0.6. However, the frequency of $\gamma$-ray BL\,Lacs with redshifts measured with emission lines is quite low \citep[$\sim$10\%, see][]{Shaw_2013:zbllacs}, which implies that we do not expect to be often in this case.

Therefore, for these objects, we expect that the UVB and VIS arms will, in general, be more effective for the detection of spectral features.
The NIR arm in this work mainly allowed a much more complete modelling of the respective contributions of the jet and host galaxy.

\subsection{Comparison with previous results}

Of the eight blazars we observed, we have measured the redshift for five objects, while for two sources we have detected absorption systems which strongly constrain the redshift. We have therefore measured (constrained) the redshift for 5/8 (7/8) of our sample compared to 34$\%$ (59$\%$) in \citet{Shaw_2013:zbllacs} for the whole sample. Though the comparison could be difficult because of the smallness of our sample, we note that \citet{Shaw_2013:zbllacs} observed five of the objects we studied (KUV\,0033--1921, BZB\,J0238--316, BZB\,J0505+0415, BZB\,J0543--5532, and BZB\,J0816--311), but they could determine a redshift for only one of them and detected no absorption system, whereas we measured three redshifts and detected four absorption systems.

\citet{Shaw_2013:zbllacs} observed the different sources in their sample for about 30 minutes with a resolution between 500 and 1000. Given the telescopes and instruments they used, their sensitivity in flux should be slightly lower than ours. The difference between our results and theirs is mainly due to the spectral resolution of X-shooter, which is a factor of 5--15 times better. Indeed, a test on BZB\,J0543--5322 and  KUV\,00311--1938, which consists in degrading the X-shooter spectra to the resolution corresponding to the \citet{Shaw_2013:zbllacs} observations (respectively with EFOSC2, resolution {$\mathcal R$} $\sim$ 350, and with LRIS at Palomar, resolution {$\mathcal R$} $\sim$ 600 -- 1000) shows that the features detected by X-shooter (Ca II H\&K for BZB\,J0543--5322 and the Mg II absorbing system for KUV\,00311--1938) are barely visible in the degraded spectra, thus justifying the non-detection with low resolution spectrographs mounted on four-meter class telescopes.

\subsection{Properties of the host galaxies}

We estimated the absolute magnitudes of the detected host galaxies and upper limits for the non-detected ones. In the case of PKS\,0301--243 and BZB\,J0505+0415, we directly used the effective radius measured by \citet{Falomo_2000:BLLac} and \citet{Nilsson_2003:BLLacs}. Under these hypotheses, our measured magnitude of the two host galaxies are compatible within the errors with the photometric ones. Therefore, we estimate that the errors on the absolute magnitude are the same as the errors on the host galaxy apparent magnitude given in Table~\ref{TabResults}.

For the detected galaxies with no previous photometric profile, we estimated the slit losses assuming that the host galaxy has an effective radius r$_e$=10 kpc with a de Vaucouleurs profile with Sersic index 1/4 \citep{Shaw_2013:zbllacs}. Under these hypotheses the slit loss corrections are between 0.64 and 0.78 magnitudes, while the corrections are $\sim$0.2 magnitudes larger using r$_e$=20 kpc.

The K-corrections were computed using the template spectrum, and for simplicity, no evolution correction was applied. The resulting absolute magnitudes can be found in Table~\ref{TabResults}; the detected galaxies are all very luminous with an average magnitude of $M_\mathrm{R} = -23.0$, marginally brighter than the average value of $\gamma$-ray blazars observed in \citet{Shaw_2013:zbllacs} and compatible with the average magnitudes of \citet{Sbaruf_2005:imared}. 

For PKS\,0447--439, KUV\,00311--1938, and BZB\,J0816--1311, we estimated a lower limit to the redshift assuming the average absolute magnitude of \citet{Shaw_2013:zbllacs} (see Table~\ref{TabResultsLL}). The redshift limit for PKS\,0447--439 is in good agreement with the values proposed by \citet{Perlman_1998:PKS0447},  \citet{Landt_2008} and \citet{Prandini_2012:PKS0447}. The limits for KUV\,00311--1938 and BZB\,J0816--1311 are, respectively, less and more constraining than the spectroscopic ones.

Emission lines were detected in PKS\,0301--243 and BZB\,J0238--3118. To investigate the environment in which they formed, we compute ([NII] $\lambda$ 6583\,\AA)/H$\alpha$ and [OIII]/H$\beta$ \citep{Baldwin_1981,Kewley_2006} line ratios under the assumption of narrow lines. The line EW measurements are used as we cannot properly evaluate the uncertainties of our modelling. This choice is justified because the lines are not very distant in wavelength and the continuum of the host galaxies does not vary much between them ($\sim $20 \%). Moreover, we checked that the computed ratios using the model are very similar to the ones computed with the equivalent widths.

We assume that the Balmer decrement is 3.1, as in case B recombination at temperature T=10$^4$ K and electron density n$_e$ $\sim$ 10$^2$-10$^4$ cm$^{-3}$ \citep{Osterbrock_2006}. This assumption is consistent with our non-detection of H$\beta$. Under these assumptions, ([NII] $\lambda$ 6583\,\AA)/H$\alpha$ $\sim$ 0.12 $\pm$ 0.1 for the total spectrum and $\sim -0.43\pm$ 0.1 for the positive spectrum of PKS\,0301--243, while [OIII]/H$\beta$ is $\sim$ 0.53 $\pm$ 0.15 for the total spectrum and $\sim$ 0.5 $\pm$ 0.08 for the positive spectrum. Notably, the first value is indicative of an AGN environment, while the second is more typical of a transition/star forming region. This is coherent with the expectation that the central region is more likely to be dominated by the AGN, while the external region is dominated by star formation, perhaps triggered by interaction with the nearby small galaxies visible in the images of \citet{Falomo_2000:BLLac}. On the other hand, ([NII] $\lambda$ 6583\,\AA)/H$\alpha$ $\ge$ 0.5 for BZB\,J0238-3118, indicating an extreme AGN environment. As [OIII] $\lambda$ 5007\,\AA\ was not detected, we cannot go further in this case.

\subsection{Absorption systems}

We detected four Mg II absorbing systems in our sample with their rest equivalent widths (REW) being 0.17\,\AA, 0.86\,\AA, 0.49\,\AA, and 0.12\,\AA. We computed the redshift path $\Delta z$ (the redshift length where we can detect Mg II systems) only for the UVB and VIS arms, as we do not expect that our sources have redshifts $\ge$ 2.75 (such that Mg II absorbers would lie in the infrared band). The covered redshift path $\Delta z$ cannot be computed exactly, because we have three objects without redshift. From the redshifts and upper limits determined, $\Delta z \ge$ 1.39 is the minimum value, while we obtain $\Delta z$ = 3.47 as a maximum possible value assuming $z$ = 1 for the three undetermined redshifts. 

The incidence of absorbing systems along the line of sight, (N/$\Delta z$),  is therefore between 2.89 $\pm$ 1.1 and 1.15 $\pm$ 0.4 at $<z>$ $\sim$ 0.31 and 0.57, respectively. We cannot directly compare this to most complete surveys of Mg II systems \citep[e.g.][]{Zhu_Menard_2013}, which are based on SDSS, because they are limited, on the one hand, to redshift $\ge$ 0.35, while, on the other hand, they do not have enough sensitivity to be compared to our data.
 
Recently, \citet{Evans_2013} have computed the incidence of Mg II absorbing systems in the redshift range 0.1 to 2.6 with a sensitivity down to 0.05\,\AA. They divide the sample into weak (0.01\,\AA $< REW<$ 0.3\,\AA), intermediate (0.3\AA $< REW <$ 1\,\AA), and strong (REW $>$ 1\,\AA) systems. The expected incidence at $z \le 1$ are about 0.74, 0.7, and 0.5, respectively (see their Figure 3). Our incidences are between 1.45 $\pm$ 0.5 and 0.57 $\pm$ 0.2 for weak and intermediate systems, respectively, and 0 for strong systems. Given the smallness of our sample, we conclude that our results are compatible with theirs. We note that X-shooter archival observation of bright blazars may be relevant to the study of weak absorbers at low redshift.

\section{Conclusion}
\label{Conclusion}

We observed eight BL\,Lac objects detected in the HE-VHE range with the \textit{Fermi} satellite and/or by ground-based Cherenkov telescopes with the X-shooter spectrograph at VLT. The goal of the campaign was to measure or constrain the redshift of our targets and to investigate the properties of the host galaxies. With our observations, we have obtained the following main results:

\begin{itemize}
\item Measurement of five redshifts and two strict lower limits from absorption systems. All redshifts values are greater than 0.2. In particular, a firm lower-limit is given for the first time at $z>0.506$ for KUV\,00311--1938 which is considered as one of the farthest sources detected at VHE, but no evidence is seen for the Ca II H\&K feature found by \citet{Piranomonte_2007:SedentarySurvey} at $z = 0.61$. Moreover, we refuted the redshift $z = 0.046$ of BZB\,J0816--1311 proposed by \citet{Jones_2009:6dFGalaxySurvey} and found a firm lower limit at $z\ge$0.288.
\item Refutation of the redshift lower limit $z\ge$1.246 proposed by \cite{Landt_2012} for PKS\,0447--439, the only source for which no extragalactic spectral feature were found in our observations. We constrain the redshift range to $0.175<z<1.51$.  
\item Measurement of the magnitudes of the host galaxies, which are marginally brighter than the average values for the host galaxies of blazars.
\item Demonstration of the possibility to determine a spectroscopic redshift for a 5500 \AA\ rest frame jet/galaxy ratio as high as 25.
\item Rough estimation of the overall properties of the gas in the two host galaxies displaying detectable emission lines.
\end{itemize}

In this work, we selected a sample of eight HE-VHE emitting BL\,Lac objects with unknown redshift, and we were highly successful in measuring their redshift or setting a lower limit to it using the X-shooter spectrograph. Comparing our results to the ones of \citet{Shaw_2013:zbllacs}, we remark that our efficiency in constraining the blazar redshifts is comparatively high, and we tentatively ascribe this effect mainly to our higher spectral resolution. All the seven BL\,Lacs with constrained redshift have a redshift greater than 0.2. The VHE emitting BL\,Lacs at those redshifts are particularly rare because the $\gamma-$ray emission, on the one hand, is absorbed by the EBL, and on the other hand, the host galaxy emission becomes more difficult to detect at greater distances. They are, however, of great interest because they put strong constraints on the EBL density \citep{Ackermann_2012:EBL,HESS_2013:EBL}.

In the near future, it is expected that the number of VHE BL\,Lac objects, with the operations of H.E.S.S. 2 and CTA, will grow substantially. We argue that X-shooter observations of their optical counterparts will be very effective in constraining their redshifts. This will, in turn, help in tuning their emission models and putting stronger constraints on the EBL.

\begin{acknowledgements}
We wish to thank the ATOM team for providing useful information on the monitoring of some of the sources in our sample and Dr. Yvonne Becherini for her contributions in the selection of the targets observed in this work.
S. Pita wishes to thank the ESO staff and in particular Christophe Martayan for their help in performing the observations.
\end{acknowledgements}

\bibliographystyle{aa} 
\bibliography{xsh_zbll}        

\end{document}